\documentclass[sigconf,nonacm]{acmart}

\newcommand\vldbdoi{XX.XX/XXX.XX}
\newcommand\vldbpages{XXX-XXX}
\newcommand\vldbvolume{19}
\newcommand\vldbissue{8}
\newcommand\vldbyear{2026}
\newcommand\vldbauthors{\authors}
\newcommand\vldbtitle{\shorttitle} 
\newcommand\vldbavailabilityurl{https://github.com/SJTU-Liquid/gMatch}
\newcommand\vldbpagestyle{empty}

\usepackage{balance}

\usepackage{svg}
\usepackage{subcaption}
\usepackage[linesnumbered,ruled,vlined]{algorithm2e}
\SetKw{Continue}{continue}
\SetKw{Break}{break}
\usepackage{mathrsfs}
\usepackage{rotating}
\usepackage{multirow}
\usepackage{threeparttable}
\usepackage{booktabs}
\usepackage{caption}
\usepackage{makecell}
\usepackage{graphicx}     
\usepackage{placeins}

\usepackage{xcolor}
\newcommand{\textph}{{\color{red}[XXXXXX]}}
\renewcommand{\textph}{gMatch}

\newcommand{\sun}[1]{{\textcolor{black}{#1}}}

\begin{document}
\title{gMatch: Fine-Grained and Hardware-Efficient Subgraph Matching on GPUs}

\author{Weitian Chen}
\affiliation{
  \institution{Shanghai Jiao Tong University}
  \city{}
  \country{}
}
\email{chenweitian@sjtu.edu.cn}

\author{Shixuan Sun}
\affiliation{
  \institution{Shanghai Jiao Tong University}
  \city{}
  \country{}
}
\email{sunshixuan@sjtu.edu.cn}

\author{Cheng Chen}
\affiliation{
  \institution{ByteDance Inc}
  \city{}
  \country{}
}
\email{chencheng.sg@bytedance.com}

\author{Yongmin Hu}
\affiliation{
  \institution{ByteDance Inc}
  \city{}
  \country{}
}
\email{huyongmin@bytedance.com}

\author{Yingqian Hu}
\affiliation{
  \institution{ByteDance Inc}
  \city{}
  \country{}
}
\email{huyingqian@bytedance.com}

\author{Minyi Guo}
\affiliation{
  \institution{Guizhou University}
  \city{}
  \country{}
}
\email{myguo@gzu.edu.cn}

\begin{abstract}
Subgraph matching is a core operation in graph analytics, supporting a broad spectrum of applications from social network analysis to bioinformatics. Recent GPU-based approaches accelerate subgraph matching by leveraging parallelism but rely on a coarse-grained execution model that suffers from scalability and efficiency issues due to high memory overhead and thread underutilization. In this paper, we propose \textph{}, a hardware-efficient subgraph matching approach on GPUs. \textph{} introduces a fine-grained execution model that reduces memory consumption and enables flexible task scheduling among threads. We further design warp-level batch exploration and lightweight load balancing to improve execution efficiency and scalability. \sun{Experiments on diverse workloads and real-world datasets show that \textph{} outperforms state-of-the-art subgraph matching methods, including STMatch, T-DFS, and EGSM, in both performance and scalability. We also compare against state-of-the-art systems for mining small patterns, such as BEEP and G$^2$Miner. While these systems achieve better performance on small datasets, \textph{} scales to substantially larger queries and datasets, where existing approaches degrade or fail to complete.}

\end{abstract}

\maketitle

\pagestyle{\vldbpagestyle}
\begingroup\small\noindent\raggedright\textbf{PVLDB Reference Format:}\\
\vldbauthors. \vldbtitle. PVLDB, \vldbvolume(\vldbissue): \vldbpages, \vldbyear.\\
\href{https://doi.org/\vldbdoi}{doi:\vldbdoi}
\endgroup
\begingroup
\renewcommand\thefootnote{}\footnote{\noindent
This work is licensed under the Creative Commons BY-NC-ND 4.0 International License. Visit \url{https://creativecommons.org/licenses/by-nc-nd/4.0/} to view a copy of this license. For any use beyond those covered by this license, obtain permission by emailing \href{mailto:info@vldb.org}{info@vldb.org}. Copyright is held by the owner/author(s). Publication rights licensed to the VLDB Endowment. \\
\raggedright Proceedings of the VLDB Endowment, Vol. \vldbvolume, No. \vldbissue\ %
ISSN 2150-8097. \\
\href{https://doi.org/\vldbdoi}{doi:\vldbdoi} \\
}\addtocounter{footnote}{-1}\endgroup

\ifdefempty{\vldbavailabilityurl}{}{
\vspace{.3cm}
\begingroup\small\noindent\raggedright\textbf{PVLDB Artifact Availability:}\\
The source code, data, and/or other artifacts have been made available at \url{\vldbavailabilityurl}.
\endgroup
}

\section{Introduction} \label{sec:introduction}

\begin{figure}[t]
    \centering
    \begin{minipage}[t]{0.3\linewidth}
        \centering
        \includegraphics[width=0.7\linewidth]{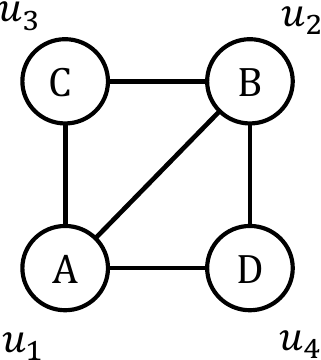}
        \subcaption{Query graph $Q$.}
        \label{fig:q}
    \end{minipage}
    \begin{minipage}[t]{0.6\linewidth}
        \centering
        \includegraphics[width=\linewidth]{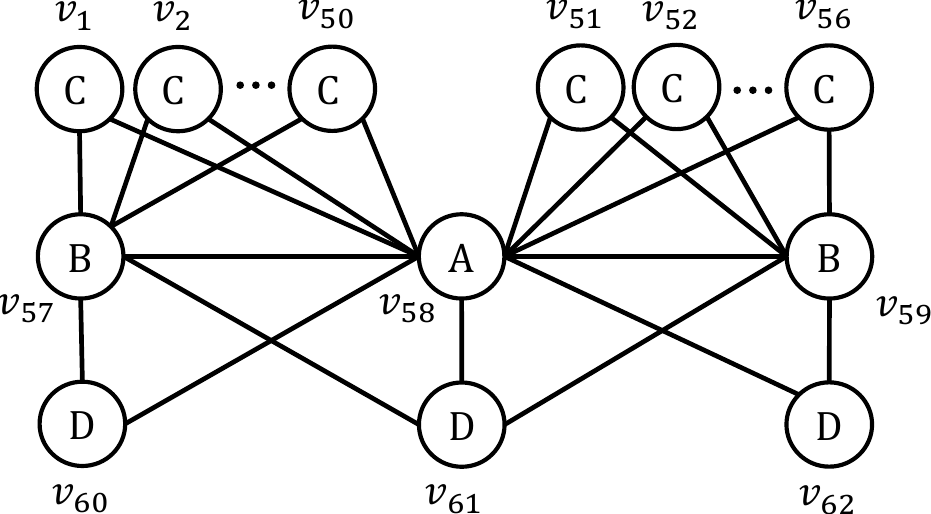}
        \subcaption{Data graph $G$.}
        \label{fig:g}
    \end{minipage}
    \caption{Example query graph and data graph.}
    \label{fig:example}
\end{figure}

Given a query graph $Q$ and a data graph $G$, \emph{subgraph matching} aims to find all embeddings of $Q$ in $G$. For example, in Figure~\ref{fig:example}, the mapping $M = \{(u_1, v_{58}), (u_2, v_{57}), (u_3, v_1), (u_4, v_{60})\}$ is a valid embedding, or \emph{match}, of $Q$ in $G$. As a fundamental operation in graph analysis, subgraph matching underpins a wide range of real-world applications, including social network recommendation~\cite{gupta2014real}, fraud detection~\cite{fraud}, bioinformatics~\cite{ri,biological}, and cheminformatics~\cite{Willett1999}.

Given the importance of subgraph matching, numerous algorithms have been proposed~\cite{rapidmatch,Harmonizing,Cartesian}. Although they adopt different optimization techniques, most follow the same core idea: iteratively extend partial matches by mapping query vertices in $Q$ to data vertices in $G$ according to a predefined order. However, since subgraph isomorphism is NP-complete, the search space can be prohibitively large. To accelerate the computation, recent studies leverage GPUs to parallelize the matching process~\cite{gpsm,gsi,cuts,egsm,stmatch,tdfs}.

All these methods adopt a \emph{coarse-grained parallel execution model} for subgraph matching, where each partial match $M$ is treated as a parallel task and a warp, the basic scheduling unit on GPUs, serves as the worker. The task involves extending $M$ by mapping the next query vertex to valid data vertices. Based on their search strategies, these methods can be classified into two categories: breadth-first search (BFS)-based and depth-first search (DFS)-based.

Early approaches such as GpSM~\cite{gpsm}, GSI~\cite{gsi}, and cuTS~\cite{cuts} adopt a BFS-based search strategy, which explores the search space level by level. At each iteration, all partial matches are extended by mapping the next query vertex to candidate data vertices. This strategy enables a large number of parallel tasks, allowing efficient GPU utilization and balanced workload distribution across warps. However, it requires storing all partial matches at every level, leading to high memory consumption due to the large search space. As a result, the scalability of BFS-based methods is severely limited by GPU memory capacity.

To reduce memory overhead, recent methods such as T-DFS~\cite{tdfs}, STMatch~\cite{stmatch}, and EGSM~\cite{egsm} adopt a DFS-based search strategy. Instead of expanding all partial matches level by level, each warp recursively explores the search space. Specifically, after generating new partial matches from a given match $M$, the warp selects one to continue the search. By avoiding the need to store all partial matches at each level, this strategy significantly improves scalability compared to BFS-based approaches.

\noindent\textbf{Key Insights.} Despite its widespread adoption, \emph{the coarse-grained parallel execution model fundamentally conflicts with real-world power-law graphs.} Because each partial match has highly dynamic and skewed workloads, while GPUs adopt the SIMT execution model and a warp provides fixed and inflexible computing resources, this design introduces fundamental limitations in both scalability and computational efficiency for GPU-based subgraph matching.

First, \emph{the execution stack incurs high memory overhead.} To handle dynamic workloads, the coarse-grained method must pre-allocate a large execution stack of size $O(|V(Q)| \times d_{\max})$ for each warp, where $d_{\text{max}}$ is the maximum degree of $G$. The stack stores intermediate results during the search, e.g., feasible candidates and derived partial matches. Since $d_{\max}$ can reach millions in power-law graphs and GPUs require hundreds of active warps for full utilization, this high memory cost limits scalability and reduces concurrency.

Second, \emph{the mismatch between the fixed warp size (32 threads) and the dynamic workloads leads to severe thread underutilization.} To extend a partial match, the coarse-grained method performs set intersections over the neighbors of previously matched data vertices. Most candidate vertices have very small degrees after filtering. A warp must still execute 32 threads in lock-step, causing pervasive idle threads and wasted GPU parallelism.

These limitations are not caused by specific implementations; they stem from the inherent structure of the coarse-grained paradigm itself. Overcoming them requires fundamentally rethinking how subgraph matching should be parallelized on GPUs.

\noindent\noindent\textbf{Our Work.} In this paper, we propose \textph{}, a hardware-efficient subgraph matching approach on GPUs. \emph{To resolve the structural mismatch between the GPU execution model and the coarse-grained parallel execution, we first introduce a fine-grained parallel execution model that redefines the basic task and the basic worker.} Instead of letting a warp process one partial match, we let a single thread process the lightweight task of checking whether a query vertex can be mapped to a data vertex. This design provides two key advantages. First, it reduces the per-worker state from $O(|V(Q)| \times d_{\max})$ to $O(|V(Q)|)$, eliminating stack-induced memory bottlenecks and enabling the system to process much larger graphs while supporting more concurrent warps. Second, it decomposes partial-match processing into naturally divisible units, allowing the workload of a single partial match to be distributed across multiple warps or tasks from different partial matches to be combined within a warp.

However, scheduling such lightweight tasks at scale is challenging, because tens of thousands of threads execute concurrently and GPU threads are organized hierarchically into warps and thread blocks. \emph{To address this, we propose a warp-level batch exploration mechanism.} We merge task groups (i.e., the tasks involved in processing a partial match) into a virtual task pool and let threads within a warp pull tasks using two shared pointers that track task boundaries. This mechanism provides minimal memory overhead, high warp utilization, and efficient composability across tasks. Such batch processing is impossible under prior coarse-grained models because each partial match is an indivisible unit; processing multiple partial matches together would require storing large intermediate states, leading to prohibitive memory costs (e.g., STMatch).

Load balancing is essential for efficient subgraph matching on GPUs. Our key observation is that although the root of the search tree may offer limited parallelism, the number of partial matches expands exponentially as the search proceeds, making the workload inherently embarrassingly parallel. When the number of tasks far exceeds the number of workers, load balancing becomes naturally easy because task variability is automatically smoothed out across a large shared task pool. \emph{Building on this observation, we introduce a lightweight load balancing strategy that includes an initialization phase to generate a large pool of partial matches as initial tasks.} This creates abundant initial parallelism and makes load balancing largely self-sustaining, minimizing the need for work stealing and reducing scheduling overhead. Consequently, work stealing becomes a supplement role in \textph{} rather than the primary mechanism, as in prior methods.

We evaluate \textph{} on two representative workloads: (1) large queries on medium-sized graphs with multiple labels, and (2) small queries on large graphs with few labels. Experimental results show that \textph{} significantly outperforms state-of-the-art methods, including STMatch~\cite{stmatch}, T-DFS~\cite{tdfs}, and EGSM~\cite{egsm}, while offering better scalability. \textcolor{black}{We also compare our approach with systems designed for mining small patterns, including BEEP~\cite{BEEP} and G$^2$Miner~\cite{g2miner}. Although these systems outperform our approach on small datasets, our method scales to substantially larger queries and datasets.}

\section{Background}

In this section, we introduce the background related to this paper.

\subsection{Preliminaries}

We focus on undirected, labeled graphs $g = (V, E)$, where $V$ is the set of vertices and $E \subseteq V \times V$ is the set of edges. For a vertex $u \in V$, let $N(u)$ denote its neighbor set and $d(u) = |N(u)|$ its degree. Each vertex is associated with a label via a labeling function $L: V \rightarrow \Sigma$, where $\Sigma$ is the label set. We denote the query graph by $Q$ and the data graph by $G$. Vertices and edges in $Q$ are referred to as \emph{query vertices} and \emph{query edges}, and those in $G$ as \emph{data vertices} and \emph{data edges}. Frequently used notations are summarized in Table~\ref{tab:notation}. Definition~\ref{def:iso} introduces \emph{subgraph isomorphism}, which we refer to as a \emph{match}. The goal of \emph{subgraph matching} is to find all matches of $Q$ in $G$.

\begin{definition} \label{def:iso}
    Given graphs $g$ and $g'$, a subgraph isomorphism is an injective function $M : V(g) \rightarrow V(g')$ s.t. 1) $\forall v \in V(g), L(v) = L(M[v])$; 2) $\forall e(v,v') \in E(g), e(M[v], M[v'])\in E(g')$.
\end{definition}

Given $Q$ and $G$ in Figure~\ref{fig:example}, $\{(u_1,v_{58}),(u_2,v_{57}),(u_3,v_{1}),(u_4,v_{60})\}$ is a match. We can observe that there exist 112 distinct matches of $Q$ in $G$. The subgraph matching problem requires finding all such mappings from $Q$ to $G$.

\begin{table}[t]
\caption{Frequently used symbols and notations.}
\centering
\resizebox{\linewidth}{!}{
\begin{tabular}{llllll}
   \toprule
   \textbf{Notations} & \textbf{Descriptions} \\
   \midrule
   $g$, $Q$ and $G$ & graph, query graph and data graph \\
   $V(g)$, $E(g)$ and $\Sigma(g)$ & vertex set, edge set and label set \\
   $d(u)$, $L(u)$ and $N(u)$ & degree, label and neighbor set of $u$ \\
   $e(u,u')$ & undirected edge connecting $u$ and $u'$ \\
   $d_{\max}, d_\text{avg}$ & maximum and average degree of the graph \\
   $C^F_M(u)$ & feasible candidate set of $u$ given partial match $M$ \\
   $C^L_M(u)$ & local candidate set of $u$ given partial match $M$ \\
   $\phi$ & matching order \\
   $N_+^{\phi}(u)$ & backward neighbors of $u$ given $\phi$ \\
   \bottomrule
\end{tabular}}
\label{tab:notation}
\end{table}

\subsection{Coarse-Grained Parallel Execution}

Although existing methods apply various optimizations, they share the same core principle: given a query graph $Q$ and a data graph $G$, they iteratively extend intermediate results along a \emph{matching order} $\phi$ to find all matches. A matching order $\phi$ is a permutation of the query vertices, where $\phi[i]$ denotes the $i$-th vertex in the sequence ($1 \leqslant i \leqslant |V(Q)|$). For each query vertex $u$, let $N_+^\phi(u)$ denote the set of neighbors that appear earlier than $u$ in $\phi$. We call $N_+ ^ \phi (u)$ the backward neighbors of $u$. The matching order is required to be \emph{connected}, i.e., $\forall u \in V(Q) \bigwedge u \neq \phi[1]$, $N_+^\phi(u) \neq \emptyset$. Algorithm~\ref{algo:CG_PSM} illustrates the parallel subgraph matching process. Based on their traversal strategies, existing methods can be categorized into BFS-based and DFS-based. 

\noindent\textbf{BFS-Search.} Lines 1–7 describe a BFS-based approach. It begins by generating an initial set $\mathcal{M}$ of partial matches by mapping the first query edge (based on the matching order $\phi$) to data edges that satisfy vertex label constraints. A partial match with $i$ query vertices corresponds to a subgraph isomorphism from the induced subgraph of the first $i$ vertices in $\phi$ to $G$. Each $M \in \mathcal{M}$ is then iteratively extended by mapping the next query vertex $u$ to candidate data vertices. At each iteration, $\mathcal{M}$ is processed in parallel, with each warp handling one partial match. Lines 19–26 perform the extension by mapping $u$ to data vertices $v$ that satisfy label and connectivity constraints, i.e., $v$ is adjacent to the data vertices mapped to $N_+^\phi(u)$. The set of such candidates is denoted by $C_M^F(u)$ (i.e., the feasible candidate sets), and the extension is executed in parallel across threads within a warp, leveraging intra-warp parallelism. Once all query vertices are matched, results are reported (Line 7). GpSM~\cite{gpsm} and GSI~\cite{gsi} adopt this BFS-based strategy. 

\noindent\textbf{DFS-Search.} Storing all partial matches at each iteration incurs high memory overhead and limits scalability. To address this, recent methods such as STMatch~\cite{stmatch}, T-DFS~\cite{tdfs}, and EGSM~\cite{egsm} employ a depth-first strategy (Lines 8–11), where each warp recursively extends a single partial match by mapping one query vertex at a time (Lines 12–16). This reduces memory usage, as only the current path in the search tree is maintained.

\noindent\textbf{Abstraction.}  
The exploration process builds a search tree where each node represents a partial match $M$, and its children $\mathcal{M}_M$ are the new partial matches obtained by mapping the next query vertex $u \in \phi$ to feasible candidates in $C_M^F(u)$. The difference lies in the traversal strategy: BFS-based methods explore the tree in a breadth-first manner, while DFS-based methods use depth-first traversal. Despite this difference, both approaches adopt a \emph{coarse-grained parallel model}, where each warp is assigned a partial match as a task and extends it independently. Intra-warp parallelism is employed to compute feasible candidates for extending each partial match.

\begin{figure}[htbp]
\centering
\includegraphics[width=\linewidth]{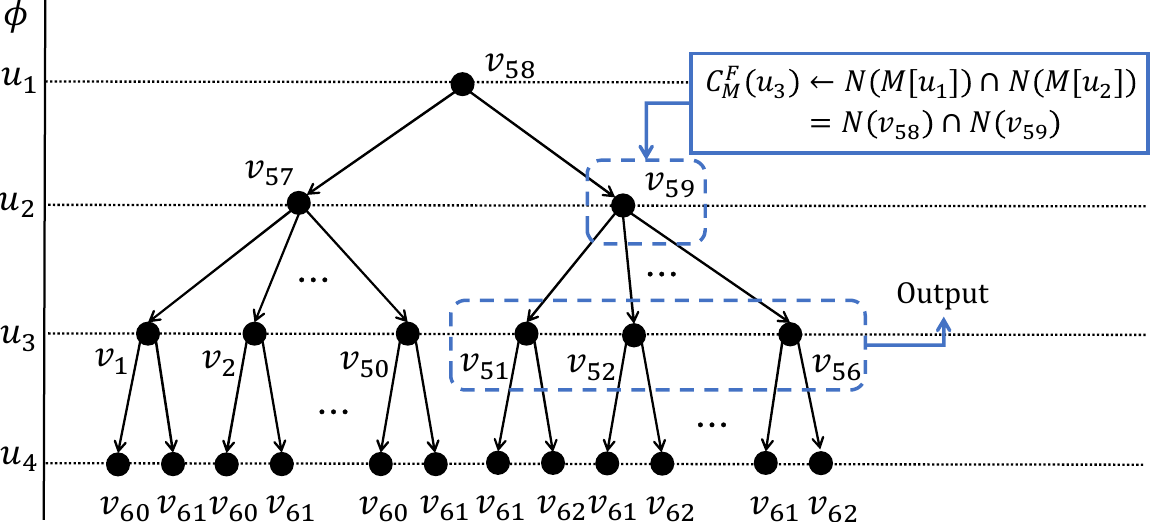}
\caption{The search tree of coarse-grained parallel execution on the example graphs in Figure~\ref{fig:example}.}
\label{fig:tree}
\end{figure}

Take $Q$ and $G$ in Figure~\ref{fig:example} as an example. The subgraph matching algorithm first generates a matching order $\phi=(u_1,u_2,u_3,u_4)$ and then performs a BFS or DFS search following $\phi$, as shown in the search tree (Figure~\ref{fig:tree}). At level $i$ of the tree, the algorithm computes feasible candidates for query vertex $\phi[i]$ based on the current partial match (i.e., the path from the root). The path $(v_{58},v_{59})$ corresponds to a match of the subgraph induced by $\{u_1,u_2\}$. To extend this match to $u_3$, the algorithm computes the intersection $N(v_{58})\cap N(v_{59})$ and retains only vertices with the same label as $u_3$. This yields the feasible candidate set $C_M^F(u_3)=\{v_{51},v_{52},...,v_{56}\}$. Note that for GPU-based subgraph matching, the set intersection operation is computed by a warp, where each thread is assigned a vertex in $N(v_{59})$ and checks the existence of this vertex in $N(v_{58})$.

\begin{algorithm}[t]
\small
\caption{Coarse-Grained Parallel Subgraph Matching}
\label{algo:CG_PSM}
\SetKwFunction{BFSSearch}{BFS-Search}
\SetKwFunction{DFSSearch}{DFS-Search}
\SetKwFunction{Search}{Search}
\SetKwFunction{Process}{Process}
\SetKwProg{proc}{Procedure}{}{}

\proc{\BFSSearch{$Q, G, \phi$}}{
    $\mathcal{M} \leftarrow \{\{(\phi[1], v), (\phi[2], v')\}| e(v, v') \in E(G)\bigwedge L(\phi[1])=L(v) \bigwedge L(\phi[2])=L(v')\}$\;
    \For{$i \leftarrow 3$ to $|\phi|$}{
        $\mathcal{M}' \leftarrow \mathcal{M}$, $\mathcal{M} \leftarrow \emptyset$\;
        \tcc{Inter-warp parallelism.}
        \text{\textbf{parallel}} \ForEach{$M \in \mathcal{M'}$}{
            $\mathcal{M} \leftarrow \mathcal{M} \bigcup$ \Process{$G, M, \phi[i]$}\;
        }
    }
    Output $\mathcal{M}$\;
}

\proc{\DFSSearch{$Q, G, \phi$}}{
    $\mathcal{M} \leftarrow \{\{(\phi[1], v), (\phi[2], v')\}| e(v, v') \in E(G) \bigwedge L(\phi[1])=L(v) \bigwedge L(\phi[2])=L(v')\}$\;
    \tcc{Inter-warp parallelism.}
    \text{\textbf{parallel}} \ForEach{$M \in \mathcal{M}$}{
        \Search{$Q, G, M, \phi, 3$}\;
    }
}

\proc{\Search{$Q, G, M, \phi, i$}}{
    \lIf{$i = |\phi| + 1$}{Output $M$, \Return}
    $\mathcal{M}_M \leftarrow$ \Process{$G, M, \phi[i]$}\;
    \ForEach{$M' \in \mathcal{M}_M$}{
        \Search{$Q, G, M', \phi, i + 1$}\;
    }
}

\proc{\Process{$G, M, u$}}{
    $C_M^F(u) \leftarrow \{\}$\;
    \tcc{Intra-warp parallelism.}
    $C_M^L(u) \leftarrow N(M[u'])$ \textit{where} $u' \in N_+ ^ \phi(u)$\;
    \textbf{parallel} \ForEach{$v \in C_M^L(u)$}{
        \If{$L(u) = L(v) \bigwedge v$ is not mapped in $M$} {
            $F \leftarrow true$\;
            \ForEach{$u'' \in N_+ ^ \phi(u)$ where $u'' \neq u'$}{
                \lIf{$v \notin N(M[u''])$}{$F \leftarrow false$, \textbf{break}}
            }
            \lIf{$F = true$}{$C_M^F(u) \leftarrow C_M^F(u) \bigcup \{v\}$}
        }
    }
    \Return $\{M \bigcup \{(u, v)\}| v \in C_M^F(u)\}$\;
}
\end{algorithm}

\noindent\textbf{Remark.} Our paper, together with prior work~\cite{study,A-Comprehensive-Survey-and-Experimental,rapidmatch}, identifies two representative workload patterns: (1) large queries over medium-sized graphs, and (2) small queries over large-scale graphs. A detailed discussion on these workloads, including why existing coarse-grained methods struggle to generalize across both efficiently, is deferred to Appendix~\ref{sec:appendix_c} due to space limitations.

\subsection{Issues in Coarse-Grained Parallel Execution}\label{sec:issues}

Although DFS-based search significantly reduces memory consumption compared to BFS-based search, we observe that its coarse-grained parallel execution leads to severe performance bottlenecks.

\noindent\textbf{Issue 1: The execution stack incurs high memory overhead, limiting both scalability and computational efficiency.} The \emph{execution stack} stores intermediate results during the search process. As shown in Algorithm~\ref{algo:CG_PSM}, when processing a partial match $M$, a warp must store both the feasible candidates $C_M^F(u)$ and the set of derived partial matches $\mathcal{M}_M$. Since the maximum size of $C_M^F(u)$ and $\mathcal{M}_M$ is bounded by the maximum degree $d_{\text{max}}$ of $G$, the coarse-grained strategy requires a buffer of size $O(d_{\text{max}})$. Dynamic memory allocation based on the actual size of $C_M^F(u)$ and $\mathcal{M}_M$ is impractical on GPUs due to high allocation overhead. Given that the DFS stack depth is $|V(Q)|$, the total space complexity of the execution stack per warp is $O(|V(Q)| \times d_{\text{max}})$.

Real-world graphs typically follow a power-law degree distribution, where a small fraction of vertices have a large number of neighbors. This leads to high memory overhead for the execution stack, limiting scalability and efficiency on GPUs, which offer abundant compute resources but limited memory.

For example, in LDBC, a social network benchmark, $d_{\text{max}}$ reaches 4 million, requiring 16 MB to store feasible candidates per partial match (assuming 4-byte vertex IDs). On an NVIDIA 4090 GPU with 128 streaming multiprocessors (SMs), each capable of 64 active warps and 24 GB memory, the execution stack quickly becomes a bottleneck. For a query graph with just five vertices and eight active warps per SM, the stack alone would exceed 80 GB of memory. Reducing the number of active warps to save memory significantly underutilizes computational resources. Additionally, the stack must reside in global memory (as it is too large to fit in shared memory), further degrading data access speed.

\begin{figure}[t]
\centering
\begin{subfigure}[b]{\linewidth}
  \includegraphics[width=\linewidth]{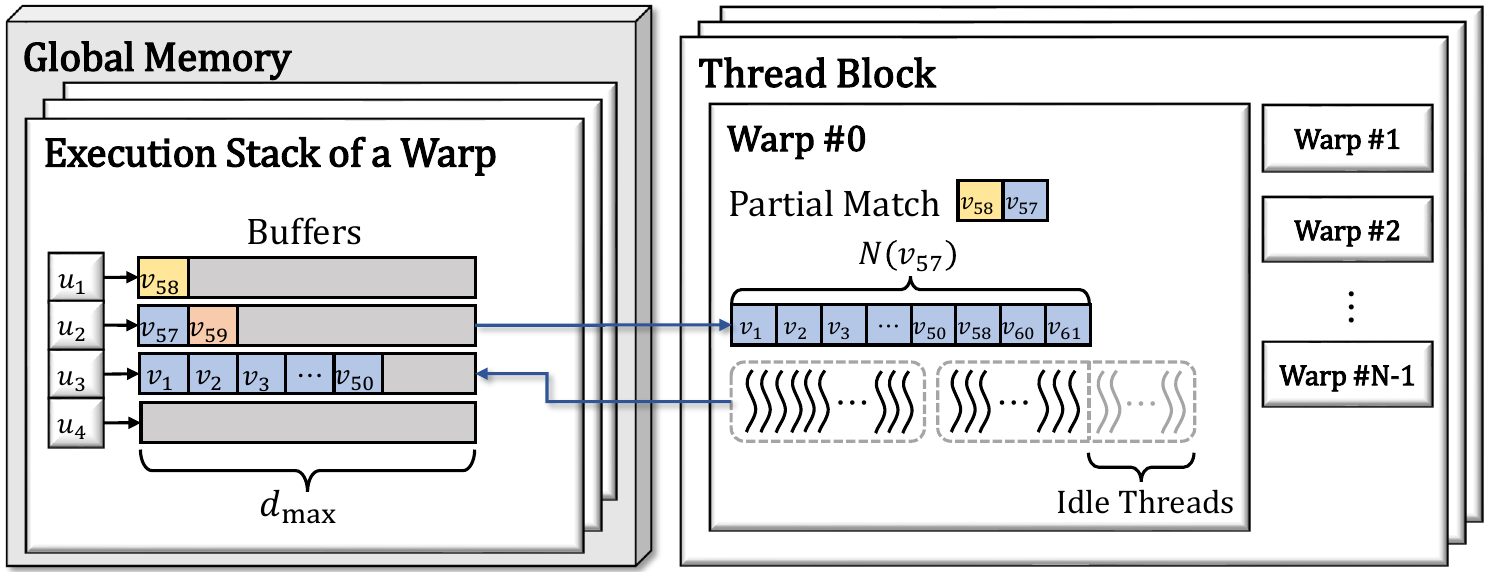}
  \caption{Extending partial match $\{(u_1,v_{58}),(u_2,v_{57})\}$.}
  \label{issue1}
\end{subfigure}
\begin{subfigure}[b]{\linewidth}
  \includegraphics[width=\linewidth]{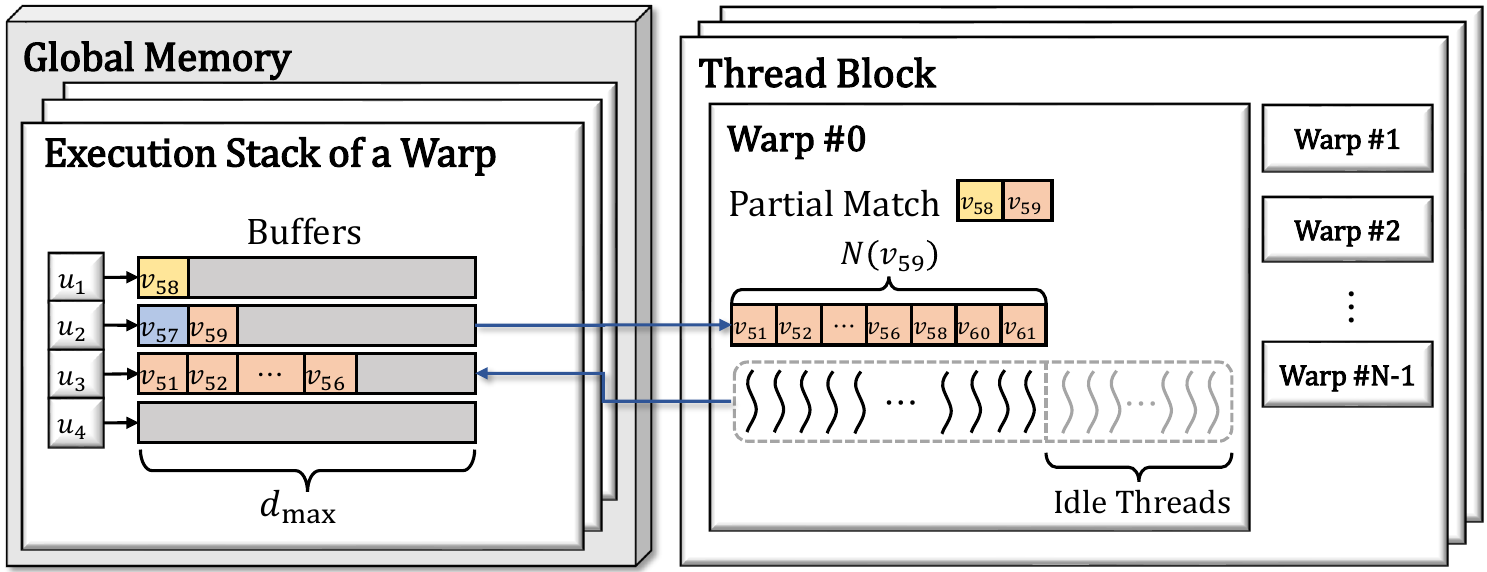}
  \caption{Extending partial match $\{(u_1,v_{58}),(u_2,v_{59})\}$.}
  \label{issue2}
\end{subfigure}
\caption{Coarse-grained parallel execution on the example graphs in Figure~\ref{fig:example}.}
\label{fig:coarse}
\end{figure}

Figure~\ref{issue1} illustrates the issue of high memory consumption of the execution stack, using the example graph of Figure~\ref{fig:example}. To extend the partial match $M=\{(u_1,v_{58}),(u_2,v_{57})\}$, the coarse-grained parallel approach computes the intersection $N(v_{58})\cap N(v_{57})$ and filters out vertices whose label is not $L(u_3)$. The result $C_M^F(u_3)=\{v_1,v_2,...,v_{50}\}$ is completely computed and materialized in the third level of the stack. The result has 50 vertices and is stored in the preallocated buffer of size $O(d_\text{max})$. This preallocated buffer incurs high memory overhead.

\noindent\textbf{Issue 2: The mismatch between the fixed warp size and the dynamic workload causes severe thread underutilization.} The coarse-grained parallel execution assigns one partial match to each warp, where 32 threads execute in lockstep. However, as shown in Lines 17–26 of Algorithm~\ref{algo:CG_PSM}, the size of the local candidate set $C_M^L(u)$ (i.e., $N(M[u'])$) varies dynamically during the search. Real-world graphs typically follow a power-law degree distribution, where most vertices have few neighbors. As a result, $|C_M^L(u)|$ is often smaller than the warp size, leading to poor intra-warp parallelism and underutilized threads.

Figure~\ref{issue2} further illustrates this issue, using the example graphs of Figure~\ref{fig:example}. When extending the partial match $M=\{(u_1,v_{58}),\allowbreak (u_2,v_{59})\}$, the coarse-grained parallel approach assigns each thread within the warp a vertex from $C_M^L(u_3)$, namely $N(v_{59})$. Each lane then check the validity of the vertex and thus construct the feasible candidates for the next level. However, there are only 9 vertices within $N(v_{59})$, and 23 threads within the warp are idle. Moreover, this underutilization problem worsens at the next level. This is because each feasible candidate identified at the current level must then verify its own local candidate set (which is also $N(v_{59})$), further amplifying the inefficiency.

STMatch~\cite{stmatch} addresses this issue using \emph{loop unrolling}, which sets a fixed unrolling factor $\sigma$ (e.g., 2, 4, or 8) to process multiple partial matches per warp by unrolling the loop at Line 15 in Algorithm~\ref{algo:CG_PSM}. While loop unrolling reduces the idle rate, it does not fully resolve thread underutilization since $\sigma$ is fixed while $|C_M^L(u)|$ varies dynamically. More critically, it increases memory consumption by expanding the execution stack to $O(|V(Q)| \times d_{\text{max}} \times \sigma)$, further limiting scalability.

\begin{table}[t]
\caption{Idle rates with varying loop unrolling sizes.}
\centering
\small
\resizebox{\linewidth}{!}{
\begin{tabular}{cccccc}
   \toprule
   \textbf{Unrolling Size} & \textbf{\textit{db}} & \textbf{\textit{en}} & \textbf{\textit{gw}} & \textbf{\textit{gh}} & \textbf{\textit{wt}}\\
   \midrule
   1         & 70.74\% & 45.14\% & 40.34\% & 48.19\% & 58.80\% \\
   2         & 50.22\% & 31.38\% & 38.62\% & 59.47\% & 29.82\% \\
   4         & 30.16\% & 20.75\% & 22.94\% & 21.62\% & 12.49\% \\
   8         & 19.32\% & 10.84\% & 15.54\% & 11.41\% &  9.80\% \\
   \bottomrule
\end{tabular}}
\label{tab:unroll}
\end{table}

To quantify this thread underutilization issue, we define the \emph{idle rate} as $\frac{1}{|\mathcal{M}|}\sum_{M \in \mathcal{M}} \frac{32 - |C_M^L(u)|}{32}$, where $\mathcal{M}$ is the set of partial matches generated during the search. We generate 100 random 12-vertex query graphs per dataset and obtain the average idle rate. Table~\ref{tab:unroll} shows the idle rate across five datasets, revealing severe hardware underutilization (see Table~\ref{datasets} for details of these datasets).

\noindent\textbf{Summary.} Existing methods suffer from significant performance limitations due to their coarse-grained parallel execution model on GPUs: (1) high memory overhead and poor scalability, as the total space complexity required for execution stacks grows as $O(|V(Q)|\times d_\text{max})$; and (2) severe thread underutilization, as the neighbor sets explored by a warp are often much smaller than the warp size, leaving many threads idle.

\section{An Overview of \textph{}}

\begin{figure}[htbp]
\centering
\includegraphics[
    width=\linewidth
]{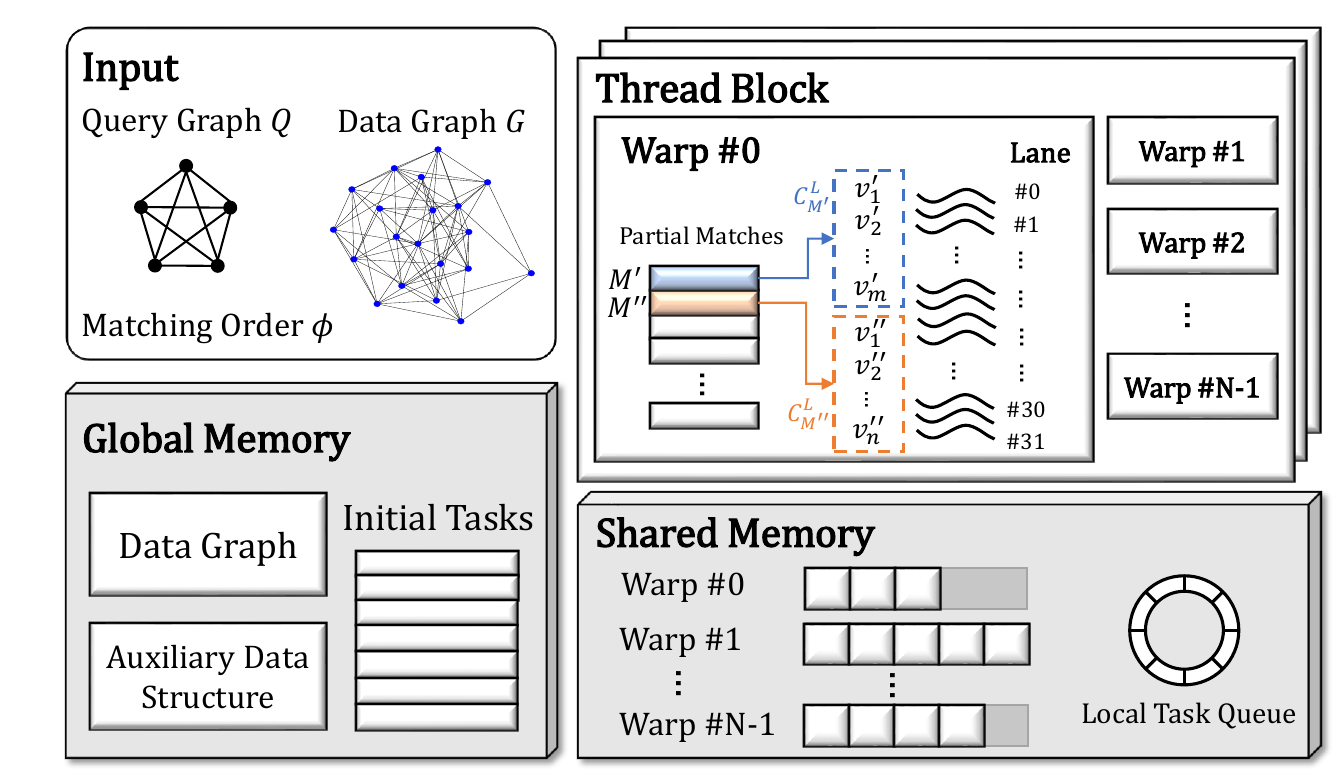}
\caption{An overview of \textph{}.}
\Description{Overview diagram of the \textph{} algorithm.}
\label{fig:overview}
\end{figure}

Figure~\ref{fig:overview} presents an overview of \textph{}, a hardware-efficient subgraph matching approach. Given $Q$ and $G$, we first generate a matching order $\phi$. As matching order selection has been extensively studied and its overhead is negligible, we generate $\phi$ on the CPU using the RI method~\cite{ri}, which has demonstrated strong performance in prior work~\cite{study,A-Comprehensive-Survey-and-Experimental}. \textph{} is compatible with other matching order generation strategies as well.

\textph{} focuses on efficient parallel execution on GPUs. Unlike prior coarse-grained parallel model that treat a warp as the basic unit of work, we propose a \emph{fine-grained parallel execution model} that treats each thread as a worker and each candidate validation as a task. This model enables fine-grained processing of partial matches and performs the search on-the-fly without materializing intermediate results, significantly reducing execution stack space and eliminating scalability constraints. Building on this model, we introduce a \emph{warp-level batch exploration} technique that groups multiple partial matches into a single warp. This aligns with the GPU architecture, where a warp is the basic scheduling unit. The technique eliminates thread underutilization without incurring additional memory overhead. To further improve scalability, we design a \emph{lightweight load balancing strategy} based on the observation that subgraph matching is an embarrassingly parallel problem. At the start of execution, we perform a BFS-based exploration to generate partial matches until a specified threshold is reached, forming an initial task pool. Warps fetch tasks independently from this pool. Once the pool is exhausted, idle warps perform work stealing to maintain balance. Thanks to the large initial task pool, stealing is rarely triggered, keeping scheduling overhead low. We detail each of these components in the following sections.

\section{Efficient Parallel Execution}

In this section, we present the efficient parallel execution strategy employed in \textph{}.

\subsection{Fine-Grained Parallel Execution}

Given $Q$, $G$, and $\phi$, existing methods \cite{stmatch,tdfs,egsm,cuts} treat each partial match $M$ as a parallel task and assign it to a warp. Suppose that $M$ contains $i$ matched vertices and $u$ is the $(i + 1)$-th query vertex in $\phi$. The warp extends $M$ by mapping $u$ to candidates $v$ in the feasible set $C_M^F(u)$. To store $C_M^F(u)$ and the resulting partial matches, each warp allocates a buffer of size $O(d_{\text{max}})$. As a result, a worker must maintain an execution stack of size $O(d_{\text{max}} \times |V(Q)|)$, leading to significant memory overhead as discussed in Section \ref{sec:issues}.

To address these limitations, we design a \emph{fine-grained execution model}. Instead of assigning a partial match $M$ to a warp as a monolithic task, we decompose it into smaller and independent tasks of the form $T_M(u, v)$,  representing an attempt to extend $M$ by mapping $u$ to a candidate vertex $v$. A task is valid if $v \in N(M[u’])$ for all $u’ \in N_+^\phi(u)$. To reduce candidate enumeration, we further restrict attention to a local candidate set $C_M^L(u) = N(M[u’])$, where $u’ = \arg\min_{u’ \in N_+^\phi(u)} |N(M[u’])|$, i.e., the neighbor set of the mapped backward query vertex with the fewest neighbors. Thus, we define the corresponding set of tasks as the task group for $M$, denoted by $T_M^G(u) = \{T_M(u, v) \mid v \in C_M^L(u)\}$.

\begin{figure}[htbp]
\centering
\includegraphics[width=\linewidth]{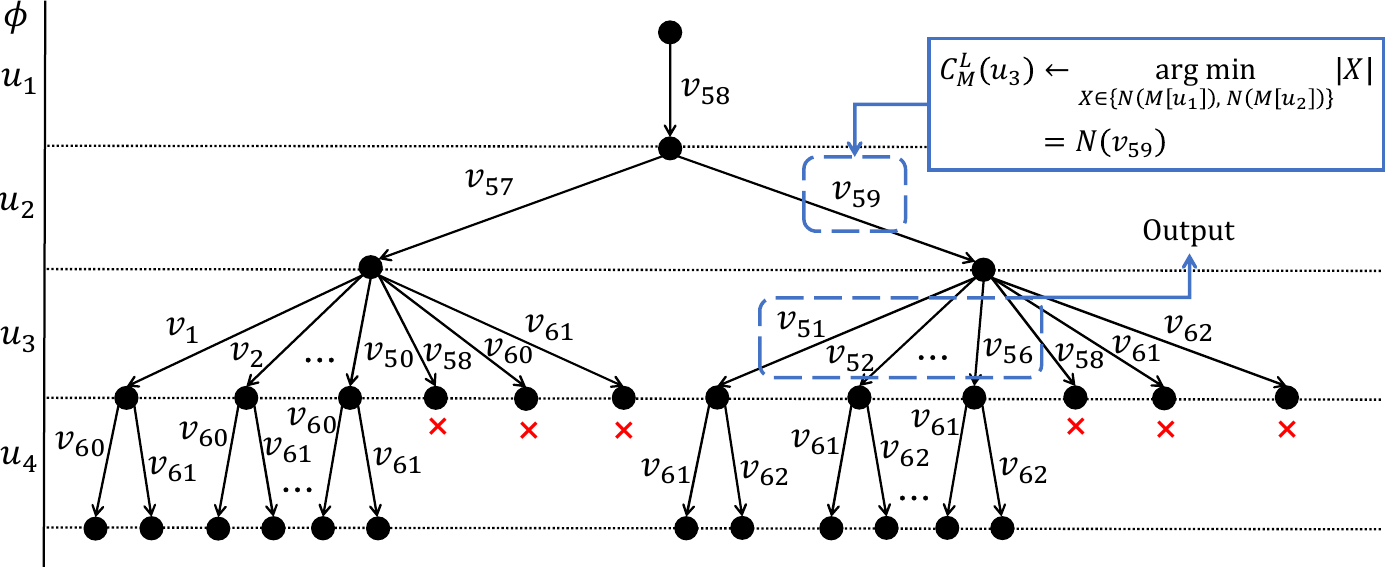}
\caption{The search tree of fine-grained parallel execution on the example graph in Figure \ref{fig:example}.}
\label{fig:fine_grained_tree}
\end{figure}

In GPUs, threads are the fundamental execution units, while warps serve as the basic scheduling units. Our execution model assigns each task group $T_M^G(u)$ to a warp, and distributes its tasks $T_M(u, v)$ to individual threads within the warp. Each thread independently processes its assigned task by verifying the feasibility of mapping $u$ to $v$. After task execution, the warp gathers the valid candidates and generates the corresponding extended partial matches. Figure~\ref{fig:fine_grained_tree} illustrates the search tree in fine-grained parallel execution, where each edge represents a task processed by a thread.

Because threads in a warp execute in lockstep, distributing tasks and gathering results incurs minimal synchronization overhead. Unlike traditional approaches storing all child matches, this fine-grained execution extends $M$ on the fly (expanding at most 32, the warp size) and avoids materializing all child matches at once and significantly reduces the space required for the execution stack.

\begin{figure}[htbp]
\centering
\includegraphics[width=\linewidth]{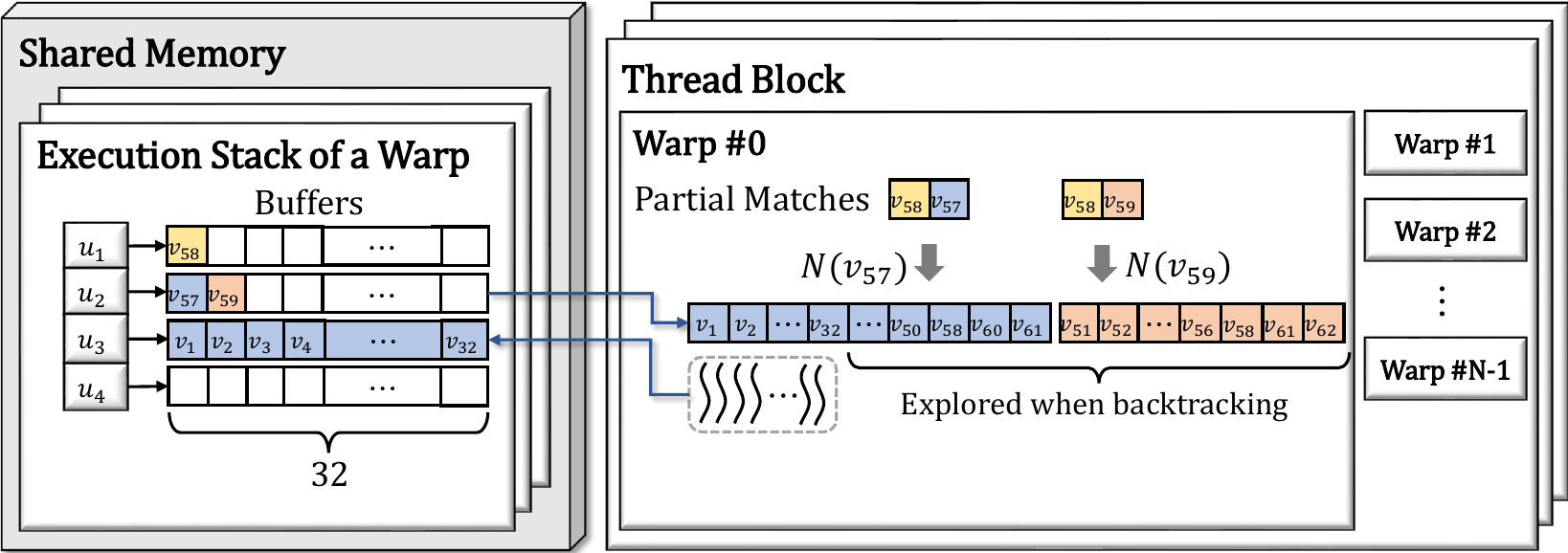}
\caption{Fine-grained parallel execution on the example graphs in Figure~\ref{fig:example}.}
\label{fig:fine_grained}
\end{figure}

Figure~\ref{fig:fine_grained} illustrates the fine-grained parallel execution model of \textph{}. We use blue to denote the subtree rooted at $v_{57}$ and red for the subtree rooted at $v_{59}$. While a warp attempts to extend the partial match $M=\{(u_1,v_{58}),(u_2,v_{57})\}$, the intersection $N(v_{57})\cap N(v_{58})$ will be computed. As $|N(v_{57})|$ is smaller than $|N(v_{58})|$, we say $C^L_M(u_3)=N(v_{57})$ and check the validity of each vertex within $C^L_M(u_3)$. Instead of computing and materializing all feasible candidates, the warp only check the validity of the first 32 vertices in $C^L_M(u_3)$. All valid candidates among them (i.e., $v_1,v_2,...,v_{32}$) are stored in the buffer of the next level. The unexplored candidates in $C^L_M(u_3)$ will be checked during backtracking.

In summary, this design aligns naturally with the GPU’s architecture and offers two key benefits: (1) it reduces space overhead, as each thread handles only one candidate and does not need to store the feasible set or generate multiple partial matches; and (2) it enables fine-grained task divisibility and flexible scheduling across threads, the fundamental execution units on GPUs.

\noindent\sun{\textbf{Remark.} Existing GPU-based methods, such as STMatch, T-DFS, and G$^2$Miner, parallelize set intersection at the warp level using SIMT execution. Specifically, each thread processes one element from a set and checks its presence in the other set via binary search. This constitutes "fine-grained" parallelism within a single intersection. However, this strategy follows a coarse-grained execution model: a set intersection is treated as an indivisible unit, and all resulting candidates must be materialized in an array. As a result, each warp is assigned to extend a single partial match and enumerate all its feasible candidates, which leads to the issues described in Section~\ref{sec:issues}. In contrast, our fine-grained computation model decomposes the extension of a partial match into independent per-candidate validation tasks and assigns each task to a single thread. This allows a warp to simultaneously process tasks from multiple partial matches, or to distribute the processing of a single partial match across multiple warps, while significantly reducing memory overhead and improving the warp utilization.}

\subsection{Warp-Level Batch Exploration}

\begin{figure}[htbp]
\centering
\includegraphics[width=\linewidth]{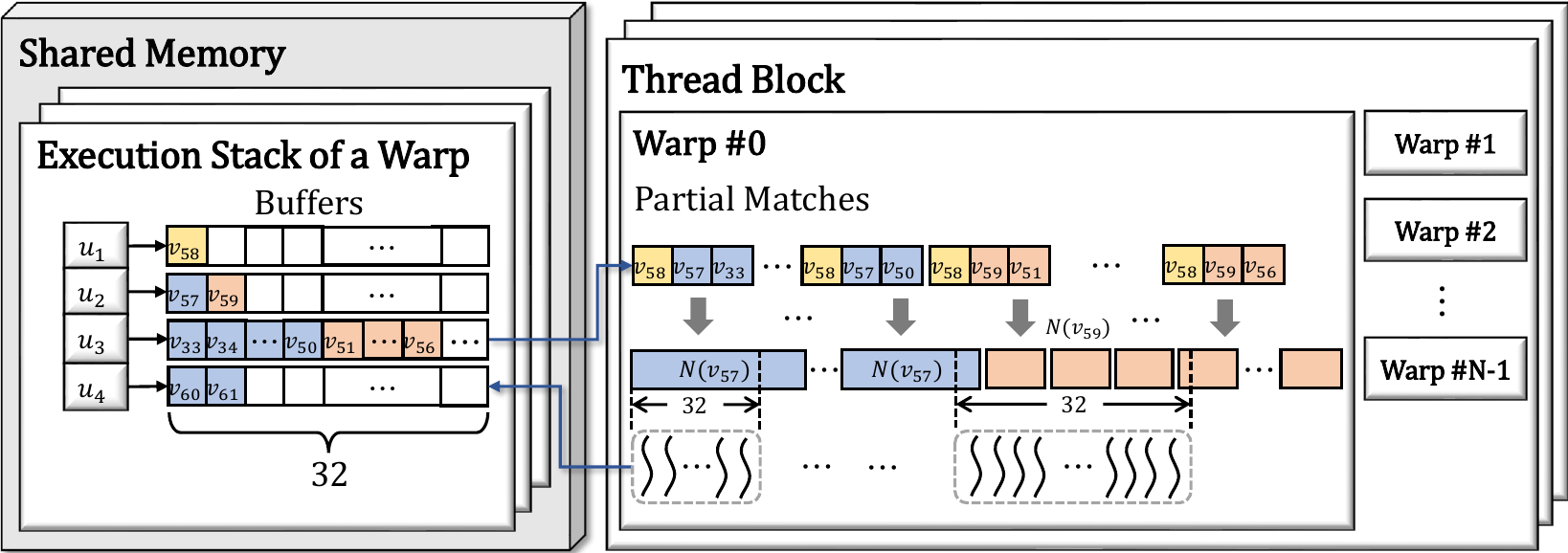}
\caption{Warp-level batch exploration.}
\label{fig:warp-batch}
\end{figure}

Real-world graphs typically follow a power-law degree distribution, where most vertices have low degree. Moreover, for labeled graphs and complex queries, advanced filtering techniques~\cite{egsm,gsi} further reduce the number of candidates. As a result, the local candidate set of a partial match is often small, and its task group cannot fully utilize all threads in a warp, leading to underutilized GPU resources. Experimental results in Section~\ref{sec:issues} confirm the severity of this issue.

To address this, we propose \emph{warp-level batch exploration} on top of the fine-grained execution model. This model enables task divisibility and flexible scheduling across threads, making it natural to process multiple task groups in a batch. Instead of assigning a single task group to each warp, we unify multiple task groups into a shared virtual task pool. Each task group $T_M^G(u)$ is defined by a partial match $M$ and its local candidate set $C_M^L(u)$. During exploration, a warp maintains up to 32 partial matches at each level, with each thread responsible for one candidate mapping. These task groups collectively form a unified task pool accessible to all threads in the warp.

Threads retrieve tasks from the pool using two lightweight pointers that track the current partial match and candidate position. This design incurs negligible memory overhead while significantly improving warp utilization without sacrificing memory efficiency. Materializing partial matches at each level can cause significant memory pressure on shared memory, with space complexity up to $O(|V(Q)|^2)$. To address this, we construct partial matches on-the-fly by traversing the execution stack from bottom to top during the computation. Since each mapping from a query vertex to a feasible candidate defines a partial match, we store, for each candidate at level $i + 1$, a pointer to its corresponding candidate at level $i$. This compact representation enables efficient reconstruction of partial matches with minimal memory overhead.

Figure~\ref{fig:warp-batch} illustrates the concept of warp-level batch exploration. This occurs during the backtracking phase of what is shown in Figure~\ref{fig:fine_grained}. To complete the tasks at level 2, the warp simultaneously processes the remaining part of $N(v_{57})$ and the entire $N(v_{59})$, treating each as a distinct batch. Since the number of unexplored candidates is fewer than 32, they can be processed within a single warp without requiring further backtracking. The feasible candidates for these two paths are stored at level 3 of the stack, denoted in blue and red, respectively. To extend these partial matches, the warp iterates through their local candidate sets using a fixed width of 32. For example, when processing the first $N(v_{57})$, the warp verifies the first 32 candidates and reports two final matches. The benefit of batch processing is more pronounced when processing $N(v_{59})$. To minimize thread idling, the warp batches candidates across multiple instances of $N(v_{59})$ together. This design inherently mitigates thread idling, as processing $N(v_{59})$ in a coarse-grained manner would lead to severe underutilization. Importantly, such batch processing is not supported by prior coarse-grained models, where each partial match is treated as an indivisible task. Processing multiple partial matches together in those models would require significant memory for storing all candidate sets and intermediate states, e.g., the loop unrolling technique in STMatch~\cite{stmatch}.

\subsection{Lightweight Load Balancing}

The search space contains an exponential number of partial matches, each forming a task group of fine-grained tasks. This leads to an embarrassingly parallel workload. In our fine-grained execution model, intra-warp load balancing is naturally achieved through a unified task pool shared by threads within a warp. We therefore focus on load balancing across warps within a thread block and across thread blocks. Warps in the same thread block can coordinate via shared memory, while communication across blocks requires global memory. To fully utilize GPU resources, we design a lightweight two-phase load balancing mechanism: (1) an initialization phase that prepopulates an initial task pool to avoid early-stage underutilization, and (2) a dynamic balancing phase that redistributes work at runtime when imbalances occur.

\noindent\textbf{Initialization Phase.}
Although the total search space is large, the number of partial matches at the root level (i.e., first-level nodes in the search tree) may be small, leading to insufficient initial parallelism. To mitigate this, we first perform a breadth-first traversal of the search tree and accumulate partial matches into an initial task pool until the number of partial matches reaches a threshold $\tau$. We then launch the fine-grained parallel execution, where each warp fetches a partial match from the pool using an atomic counter when it runs out of work. 

The parameter $\tau$ is critical because it regulates the need for dynamic load balancing. When the number of initial partial matches significantly exceeds the number of available warps (typically hundreds to thousands on modern GPUs), load balancing becomes naturally easy because task variability is automatically smoothed out across a large shared task pool, reducing the necessity for explicit load-balancing mechanisms. This principle makes it straightforward to choose $\tau$ empirically: a larger value leads to better workload balance but comes with higher memory consumption. In our experiments, we set $\tau$ to $10^6$, achieving a good balance between memory consumption and efficiency. The experiments on the choice of $\tau$ are presented in Appendix~\ref{sec:appendix_b}.

\noindent\textbf{Dynamic Balancing Phase.}
Once the initial task pool is exhausted, we shift to dynamic load redistribution. We follow the same high-level idea as STMatch~\cite{stmatch}, idle warps receive half of the work from busy warps by splitting the execution stack of the selected busy warp, but our implementation is simpler and better aligned with our two-level load balancing.

Specifically, when a warp runs out of tasks, it marks itself as idle in shared memory and enqueues its warp ID into a concurrent queue maintained per thread block. Each block also maintains a status array to track warp activity. Busy warps periodically check the queue before each recursive call. If a busy warp dequeues an idle warp ID, it splits its current task stack in half and assigns the split portion to the idle warp, updating the status accordingly. Concurrently, idle warps continuously poll their status in the array. Once an idle warp detects its status change to active, it resumes execution with the newly assigned tasks. This avoids the costly stack scanning in STMatch and reduces contention, even though the busy warp selected may not always have the largest remaining workload. To support inter-block load balancing, the same mechanism is replicated at the block level using global memory for the queue and status array. In summary, by exploiting the intrinsic parallel structure of subgraph matching, our design makes work stealing a supplementary rather than primary mechanism for achieving load balancing.

\section{Implementation and Cost Analysis}\label{sec:implementation}

In this section, we first introduce the implementation and then have a discussion on the cost of \textph{}.

\subsection{Implementation Details}

Algorithm~\ref{algo:gmatch} presents the details of \textph{}. We begin by performing a BFS to generate partial matches and populate the initial task pool, up to a threshold $\tau$. Each warp executes independently using an execution stack $S$, which is a 2D array of size $|V(Q)| \times 32$. Each element $S[i][j]$ represents the $j$-th thread’s state at depth $i$ and contains four fields:  (1) $v$: a candidate data vertex,  (2) $F$: a flag indicating whether mapping $\phi[i]$ to $v$ is valid, (3) $pid$: the index of the parent vertex in $S[i - 1]$ corresponding to $\phi[i - 1]$, (4) $C$: a pointer to the local candidate set for the current partial match.

Upon retrieving a partial match $M$ from the task pool, we initialize the execution stack $S$ and begin the search (Lines 2–4). Let $l$ denote the current recursion depth, where $\phi[l]$ is the query vertex to be extended. At Line 7, we generate the \emph{virtual task pool}. As shown in Lines 17–24, each thread traverses the execution stack to reconstruct its corresponding partial match on the fly without materializing it and computes the local candidate set for $\phi[l]$. Each resulting task group forms part of the virtual task pool, enabling fine-grained scheduling at the warp level.

Each thread in the warp then retrieves a task from the virtual task pool (Lines 9–10). To avoid synchronization overhead, a designated leading thread assigns one task to each thread in the warp (Lines 25–32). Since a warp has only 32 threads, this assignment incurs negligible cost. Each thread then processes its assigned task by checking whether the candidate data vertex can extend the partial match. This involves verifying the label constraint, ensuring the vertex has not been used, and checking the edge constraints (Lines 33–41). If the generated partial match reaches the length of the matching order, the result is output (Lines 12–13). Otherwise, we use a warp-level collective primitive to check whether any thread has a feasible candidate; if so, the search continues (Lines 15–16).

\begin{algorithm}[t]
\small
\caption{\textph{}}
\label{algo:gmatch}
\SetKwInOut{Input}{Input}
\SetKwInOut{Output}{Output}
\SetKwFunction{ScatterTask}{ScatterTask}
\SetKwFunction{Search}{Search}
\SetKwFunction{GenerateTask}{GenerateTask}
\SetKwFunction{Process}{Process}
\SetKwProg{proc}{Procedure}{}{}
\Input{Query graph $Q$, data graph $G$, matching order $\phi$, and initial task pool size $\tau$.}
\Output{All matches from $Q$ to $G$.}

$\mathcal{M} \leftarrow $ generate $\tau$ partial matches with BFS-Search\;

\tcc{Inter-warp parallelism}
\textbf{parallel} \ForEach{$M \in \mathcal{M}$}{
    Initialize execution stack $S$ of a warp with $M$\;
    \Search{$\phi, S, |M|$}\;
}

\proc{\Search{$\phi, S, l$}}{
    $i \leftarrow 0, j \leftarrow 0, k \leftarrow 0$\;
    \tcc{Each lane processes an element.}
    \GenerateTask{$\phi, S, l, LANE\_ID$}\;
    
    \While{$true$}{
        \If{$LANE\_ID = 0$}{$(i, j, k) \leftarrow$\ScatterTask{$S, l, i, j, k$}\;}
        \tcc{Each lane processes a task.}
        $S[l][LANE\_ID].F \leftarrow$\Process{$\phi, S, l, k, LANE\_ID$}\;
        
        \eIf{$l = |\phi| - 1$}{
            \lIf{$S[l][LANE\_ID].F = true$}{Output}
        }{
            $F \leftarrow \textsf{ballot\_sync}(\text{0xFFFF FFFF}, S[l][LANE\_ID].F)$\;
            \lIf{$F \neq 0$}{\Search{$\phi, S, l+1$}}
        }
    }
    
}

\proc{\GenerateTask{$\phi, S, l, lid$}}{
    $S[l][lid].C \leftarrow \emptyset$, $C \leftarrow \emptyset$, $pid \leftarrow lid$\;
    \lIf{$S[l-1][lid].F = false$}{\Return}
    
    \For{$i \leftarrow l - 1 \text{ to } 0$}{
        $v \leftarrow S[i][pid].v$, $pid \leftarrow S[i][pid].pid$\;
        \If{$\phi[i] \in N_+^\phi (\phi[l])$ and ($|N(v)| \leqslant |C|$ or $C \neq \emptyset$)}{
        $C \leftarrow N(v)$\;
        }
    }
    $S[l][lid].C \leftarrow C$\;
}

\proc{\ScatterTask{$S, l, i, j, k$}}{
    \While{$i < 32$}{
        \While{$j < |S[l][i].C|$}{
            $S[l][k].v \leftarrow S[l][i].C[j]$, $S[l][k].pid \leftarrow i$\;
            $j \leftarrow j + 1$, $k \leftarrow k + 1$\;
            \lIf{$k = 32$}{\Return $(i, j, k)$}
        }

        $i \leftarrow i + 1$, $j \leftarrow 0$\;
    }

    \Return $(i, j, k)$\;
}

\proc{\Process{$\phi, S, l, k, lid$}}{
    \lIf{$lid < k$}{\Return $false$}
    $u \leftarrow \phi[l]$, $v \leftarrow S[l][lid].v$, $pid\leftarrow S[l][lid].pid$\;
    \lIf{$L(v) \neq L(u)$}{\Return $false$}
    \For{$i \leftarrow l - 1 \text{ to } 0$}{
        $u' \leftarrow \phi[i]$, $v' \leftarrow S[i][pid].v$, $pid \leftarrow S[i][pid].pid$\;
        \If{$v = v'$ or ($u' \in N_+^\phi (u)$ and $v \notin N(v')$)}{
            \Return $false$\;
            }
    }
    \Return $true$\;
}
\end{algorithm}

\subsection{Cost Analysis}

We now analyze the space and time complexity of our parallel execution model and highlight its advantages over existing methods.

\noindent\textbf{Space Cost.}
Since all methods take the same input and produce the same output, we focus on the additional memory used during execution, specifically, the space required for the execution stack and load balancing structures. As the warp is the basic scheduling unit on GPUs, we analyze space usage at the warp level. Let $W$ be the number of threads per warp, and $B$ the number of warps per thread block. The execution stack depth equals $|V(Q)|$, and each thread maintains one candidate per level. Thus, the stack occupies $O(W \times |V(Q)|)$ space per warp. Given that $|V(Q)|$ is small (e.g., 12), the execution stack requires only a few kilobytes and comfortably fits in shared memory. As a result, shared memory usage is minimal and does not limit warp occupancy.

For load balancing, the local task queue within a thread block stores at most $B$ entries (one per warp), requiring $O(B)$ space—also small enough for shared memory. The global task queue, used for inter-block coordination, is allocated in global memory. Its size is proportional to the number of blocks (typically hundreds), which is negligible compared to the available global memory (usually tens of gigabytes).

\noindent\textbf{Time Cost.}
The total time cost is determined by the size of the search space and the efficiency of processing each partial match. Given $Q$, $G$, and $\phi$, the search space is fixed. We thus focus on analyzing the time to process a single partial match $M$.

Let $u$ be the current query vertex to match. For each candidate $v \in C_M^L(u)$, feasibility checking involves two steps: (1) verifying that $v$ is a neighbor of all data vertices mapped to $u$’s backward neighbors, i.e., for each $u’ \in N_+^\phi(u)$, we check whether $v \in N(M[u’]))$ via binary search, costing $O(\log |N(M[u’])|)$;
(2) checking whether $v$ has already been used in $M$, which requires scanning the execution stack in $O(|M|)$ time. We fuse these checks into a single loop to improve efficiency. Thus, the total cost of processing $M$ is: $O(\sum_{v \in C_M^L(u)} (|M| + \sum_{u' \in N_+ ^ \phi (u)} \log | N(M[u'])|)$.

This cost is comparable to that of existing methods, but our approach achieves significantly better hardware utilization through fine-grained parallelism and warp-level batch exploration. The overhead of the scatter operation is negligible, as a warp processes at most 32 candidates.

\noindent\textbf{Advantages over Existing Methods.}  
Compared with existing methods, our approach does not reduce the size of the search space or the time complexity of processing a partial match. Instead, it focuses on maximizing hardware utilization. First, our method reduces the space complexity of the execution stack per warp from $O(|V(Q)| \times d_{\text{max}})$ to $O(|V(Q)|)$. This removes memory-related constraints on parallel execution, ensuring that the number of concurrent warps is not limited by stack size while also freeing substantial memory to handle larger graphs. Moreover, the stack can reside entirely in shared memory, improving access efficiency. Second, warp-level batch exploration enables full utilization of computing resources within a warp without incurring significant overhead, even when task groups are small. Finally, we employ a lightweight load balancing strategy that uses an initialization phase to prepopulate tasks, reducing the need for work stealing during execution.

\section{Experiments}

In this section, we present the experimental settings and the results.

\subsection{Experimental Setup}
\label{sec:experimental_setup}

\textbf{Methods Under Study.} We compare \textbf{\textph{}} with the latest GPU-based subgraph matching algorithms, including \textbf{STMatch}~\cite{stmatch}, \textbf{EGSM}~\cite{egsm}, and \textbf{T-DFS}~\cite{tdfs}. \sun{We also include two systems designed for mining small patterns: \textbf{BEEP}~\cite{BEEP} and \textbf{G$^2$Miner}~\cite{g2miner}. Their source code is publicly available on GitHub\footnote{\url{https://github.com/HPC-Research-Lab/STMatch} (commit: f98462a)}$^,$\footnote{\url{https://github.com/RapidsAtHKUST/EGSM} (commit: de854d5)}$^,$\footnote{\url{https://github.com/lyuheng/tdfs} (commit: 05ef1b1)}$^,$\footnote{\url{https://github.com/Nagi-Research-Group/BEEP} (commit: 3eca170)}$^,$\footnote{\url{https://github.com/chenxuhao/GraphMiner} (commit: 2a76e3f)}.} To ensure fair comparison, we unify the matching order in STMatch and T-DFS by replacing their original strategies with the RI~\cite{ri} ordering, consistent with our method. \sun{For EGSM, BEEP, and G$^2$Miner, we use their default ordering, which is specifically optimized for each framework.} EGSM leverages a candidate graph for efficient candidate filtering and retrieval, which is particularly effective for workloads with large query graphs and medium-sized data graphs. We adopt the same candidate graph mechanism in \textph{} when comparing against EGSM. However, we disable this feature when comparing against STMatch, T-DFS, BEEP, and G$^2$Miner, as they do not use candidate graphs and gain minimal benefit from candidate graphs for their target workloads. Our method’s compatibility with candidate graphs highlights its generality.

\noindent\textbf{Testbed.} All competing algorithms are compiled with GCC 10.5 and NVCC 12.5. By default, we run our experiments on a Linux machine with an Intel Xeon Gold 6430 CPU and 256 GB main memory. The machine is equipped with an NVIDIA RTX 4090 GPU with 128 SMs and 24 GB global memory space.

\noindent\textbf{Datasets.} We use 11 graph datasets in our experiments, as summarized in Table~\ref{datasets}. These include eight real-world social networks (\emph{en}, \emph{gh}, \emph{gw}, \emph{wt}, \emph{lj}, \emph{pk}, \emph{fr}, and \emph{ot}), one co-authorship network (\emph{db}), and two synthetic datasets (\emph{ld} and \emph{rm}). The dataset \emph{ld} is from the LDBC social network benchmark~\cite{DBLP:conf/sigmod/ErlingALCGPPB15}. We also use PaRMAT~\cite{parmat} to generate an RMAT graph \emph{rm} because it can imitate the structure of real-world networks and allows us to control the graph scale to meet the memory constraints of a single GPU. The remaining datasets are from SNAP~\cite{snap}. We choose the datasets \emph{en}, \emph{gh}, \emph{gw}, \emph{wt}, and \emph{db} for comparison with EGSM, as they were also used in its experiments. These datasets are obtained directly from the EGSM authors' release. Additionally, we include the datasets \emph{pk}, \emph{fr}, \emph{ot}, and \emph{lj} for comparison with STMatch and T-DFS, following the same datasets used in their original experiments. We choose the dataset \emph{ld} as it is a popular benchmark for graph query and processing. 

We categorize the datasets based on their edge count: datasets with fewer edges belong to the first category, and those with more to the second. We use large query graphs on the first category and small query graphs on the second. All data graphs in the first category are labeled, where each vertex is assigned with a label uniformly at random. All datasets in the second category are unlabeled (i.e., $|\Sigma|=1$).

\begin{table}[htbp]
\caption{\textcolor{black}{The detailed statistics of datasets.}}
\centering
\resizebox{\linewidth}{!}{
\begin{tabular}{ccccccc}
   \toprule
   \textbf{Datasets} & \textbf{Abbr.} & $|V|$ & $|E|$ & $|\Sigma|$ & $d_\text{avg}$ & $d_\text{max}$ \\
   \midrule
   Enron & \textit{en} & 36,692 & 183,831 & 16 & 10.0 & 1,383 \\
   GitHub & \textit{gh} & 37,700 & 289,003 & 16 & 15.3 & 9,458 \\
   Gowalla & \textit{gw} & 196,591 & 950,327 & 16 & 9.7 & 14,730 \\
   DBLP & \textit{db} & 317,080 & 1,049,866 & 16 & 6.6 & 343 \\
   WikiTalk & \textit{wt} & 2,394,385 & 4,659,565 & 64 & 3.9 & 100,029 \\
   \midrule
   Pokec   & \textit{pk} & 1,632,803 & 22,301,964 & 1 & 27.3 & 14,854 \\
   LiveJournal & \textit{lj} & 4,847,571 & 42,851,237 & 1 & 17.7 & 14,815 \\
   Orkut & \textit{ot} & 3,072,441 & 117,185,083 & 1 & 76.3 & 33,313 \\
   LDBC  & \textit{ld} & 29,982,730 & 175,860,387 & 1 & 11.8 & 4,282,595 \\
   Friendster & \textit{fr} & 65,608,366 & 1,806,067,135 & 1 & 55.1 & 5,214 \\
   \textcolor{black}{RMAT} & \textcolor{black}{\textit{rm}} & \textcolor{black}{99,999,983} & \textcolor{black}{2,800,000,000} & \textcolor{black}{1} & \textcolor{black}{56.0} & \textcolor{black}{112,497} \\
   \bottomrule
\end{tabular}}
\label{datasets}
\end{table}

\begin{figure}[htbp]
\centering
\includegraphics[width=0.5\linewidth]{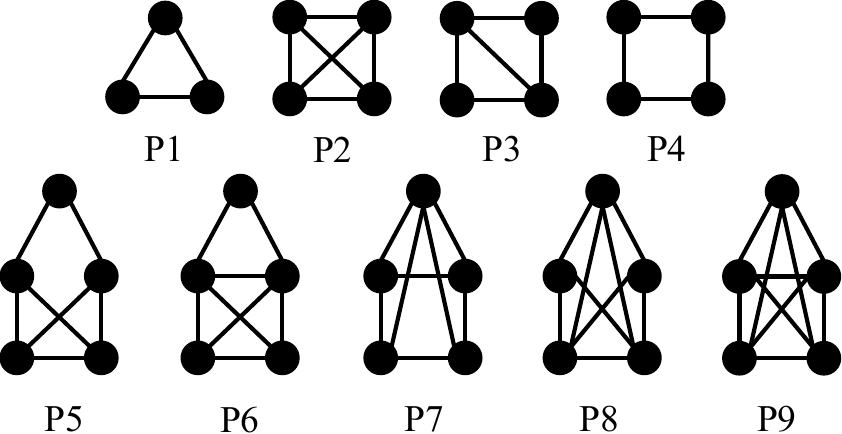}
\caption{Query graphs.}
\label{patterns}
\end{figure}

\renewcommand{\arraystretch}{0.8}

\begin{table*}[t]
    \centering
    \small
    \caption{\sun{Comparison of search time (millisecond) among T-DFS (TD), STMatch (ST), BEEP (BE), G$^2$Miner (G2), and gMatch (GM). OOT indicates timeouts, OOM indicates out-of-memory, and N/A indicates unsupported queries in the open-source code.}}
    \label{tab:g2}
    \setlength{\tabcolsep}{3.5pt}
    
    \begin{tabular}{c|ccccc|ccccc|ccccc}
        \toprule
        & \multicolumn{5}{c|}{\textit{pk}} & \multicolumn{5}{c|}{\textit{lj}} & \multicolumn{5}{c}{\textit{ot}} \\
        \cline{2-16}
         & TD & ST & BE & G2 & GM & TD & ST & BE & G2 & GM & TD & ST & BE & G2 & GM \\
        \midrule
        P1 & 84 & 9,030 & 129 & \textbf{12} & {23} & 112 & 12,413 & 239 & 87 & \textbf{31} & 463 & 12,486 & 661 & \textbf{92} & 122 \\
        P2 & 431 & 8,566 & 172 & \textbf{22} & {86} & 2,083 & 4,344 & 350 & \textbf{116} & 422 & 5,601 & 9,587 & 840 & \textbf{190} & 1,077 \\
        P3 & 987 & 3,667 & 229 & \textbf{32} & {323} & 5,665 & 11,991 & 562 & \textbf{118} & {2,080} & 19,244 & 43,329 & 1,380 & \textbf{299} & {7,661} \\
        P4 & 5,473 & 15,134 & N/A & \textbf{336} & {1,626} & 12,932 & 29,071 & N/A & \textbf{885} & {4,127} & OOT & 285,055 & N/A & \textbf{9,730} & {47,113} \\
        P5 & 13,429 & 12,549 & N/A & N/A & \textbf{4,390} & 926,807 & 641,476 & N/A & N/A & \textbf{334,373} & 1,234,177 & 4,153,864 & N/A & N/A & \textbf{526,114} \\
        P6 & 29,846 & 10,260 & \textbf{773} & N/A & {7,935} & 3,865,506 & 509,852 & \textbf{16,875} & N/A & {637,667} & 2,568,786 & 1,449,627 & \textbf{9,600} & N/A & {807,067} \\
        P7 & 13,225 & 5,916 & 6,910 & N/A & \textbf{3,295} & 842,156 & 259,700 & \textbf{23,703} & N/A & {305,968} & 976,353 & 848,283 & \textbf{134,389} & N/A & {322,477} \\
        P8 & 6,223 & 4,569 & \textbf{587} & N/A & {1,646} & 520,329 & 153,237 & \textbf{12,218} & N/A & {221,493} & 219,505 & 267,167 & \textbf{6,268} & N/A & {104,190} \\
        P9 & 1,754 & 8,177 & 346 & \textbf{55} & {288} & 106,059 & 20,799 & \textbf{3,995} & {4,347} & 25,114 & 46,943 & 36,807 & 2,832 & \textbf{825} & {7,962} \\
        
        \midrule
        
        & \multicolumn{5}{c|}{\textit{ld}} & \multicolumn{5}{c|}{\textit{fr}} & \multicolumn{5}{c}{\textit{rm}} \\
        \cline{2-16}
         & TD & ST & BE & G2 & GM & TD & ST & BE & G2 & GM & TD & ST & BE & G2 & GM \\
        \midrule
        P1 & 571 & OOM & OOM & \textbf{45} & 172 & OOM & OOM & OOM & \textbf{1,224} & 4,354 & OOM & OOM & OOM & \textbf{2,551} & 5,714 \\
        P2 & 2,149 & OOM & OOM & \textbf{83} & 549 & OOM & OOM & OOM & \textbf{4,090} & {12,132} & OOM & OOM & OOM & \textbf{3,784} & 5,866 \\
        P3 & 12,387 & OOM & OOM & OOM & \textbf{5,844} & OOM & OOM & OOM & OOT & \textbf{53,556} & OOM & OOM & OOM & OOM & \textbf{51,243} \\
        P4 & 377,456 & OOM & N/A & 205,991 & \textbf{152,834} & OOM & OOM & N/A & OOT & \textbf{2,806,445} & OOM & OOM & N/A & OOT & \textbf{3,033,834} \\
        P5 & 315,378 & OOM & N/A & N/A & \textbf{149,406} & OOM & OOM & N/A & N/A & \textbf{1,068,031} & OOM & OOM & N/A & N/A & \textbf{266,772} \\
        P6 & 439,141 & OOM & OOM & N/A & \textbf{167,778} & OOM & OOM & OOM & N/A & \textbf{1,519,540} & OOM & OOM & OOM & N/A & \textbf{225,545} \\
        P7 & 153,001 & OOM & OOM & N/A & \textbf{95,427} & OOM & OOM & OOM & N/A & \textbf{731,884} & OOM & OOM & OOM & N/A & \textbf{138,211} \\
        P8 & 19,294 & OOM & OOM & N/A & \textbf{10,191} & OOM & OOM & OOM & N/A & \textbf{235,798} & OOM & OOM & OOM & N/A & \textbf{42,032} \\
        P9 & 3,828 & OOM & OOM & \textbf{114} & {910} & OOM & OOM & OOM & \textbf{4,643} & {33,776} & OOM & OOM & OOM & \textbf{3,801} & 6,108 \\
        \bottomrule
    \end{tabular}
\end{table*}

\begin{figure*}[t]
\centering
\includegraphics[width=\textwidth]{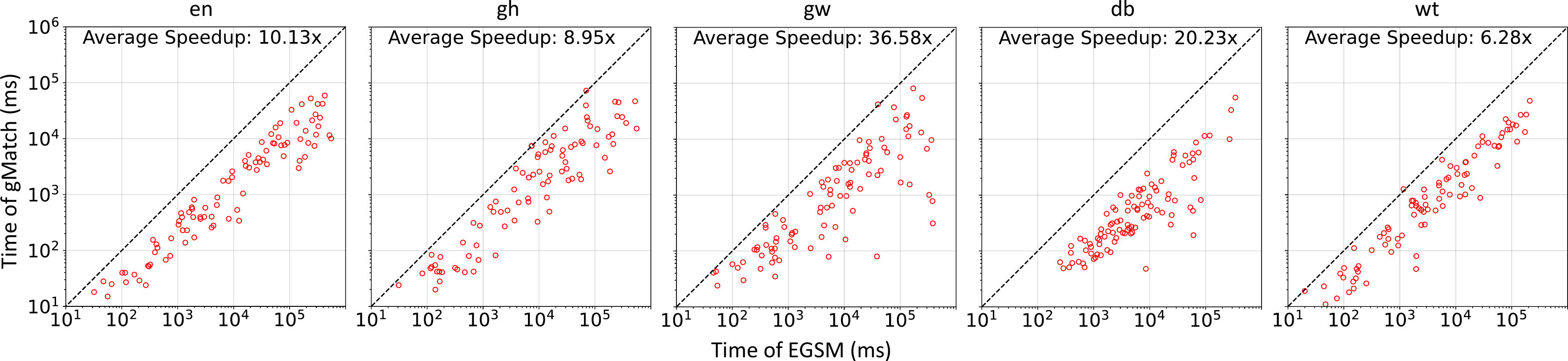}
\caption{Comparison of search time between EGSM and \textph{} (millisecond).}
\label{fig:egsm_comparison}
\end{figure*}

\renewcommand{\arraystretch}{1}

\noindent\textbf{Query Sets.} We evaluate the performance of subgraph matching algorithms across different query graph sizes. For small query graphs ($|V(Q)|\leq5$), we use the predefined query graphs shown in Figure~\ref{patterns}. We consider 9 patterns which are commonly used in existing work \cite{tdfs,Peregrine,g2miner,rapidmatch} for performance evaluation. All of them are unlabeled and will be used for testing on unlabeled data graphs. For larger query graphs, we generate 100 random queries per dataset using the same procedure as in EGSM and other prior work \cite{Harmonizing,rapidmatch,Cartesian}. Specifically, to generate an $n$-vertex query graph, we start with a random seed vertex from the data graph and iteratively expand it by adding random neighboring vertices, along with all their connecting edges to the existing subgraph, until the target size of $n$ vertices is reached.

\noindent\textbf{Metrics.} The total query time $t_{\text{query}}$ for a query is divided into filtering time $t_{\text{filtering}}$, data transfer time ($t_{\text{transfer}}$), and search time ($t_{\text{search}}$). We define $t_{\text{query}}$ as the time from when $G$ and $Q$ are loaded into host memory until all results are found. $t_{\text{filtering}}$ measures the time for candidate filtering and auxiliary structure construction, $t_{\text{transfer}}$ captures the time to move $G$, $Q$, and auxiliary data to GPU memory, and $t_{\text{search}}$ accounts for the time to enumerate all matches. To ensure fair comparison, we report only $t_{\text{search}}$. This is because, for large queries, \textph{} and EGSM share the same auxiliary structures, so any difference lies solely in $t_{\text{search}}$. For small queries, STMatch performs degree-based filtering on the CPU, which becomes a bottleneck on large datasets (e.g., \emph{ot}), whereas \textph{} and T-DFS implement the same filtering on the GPU, incurring negligible cost. Therefore, $t_{\text{search}}$ provides a fair metric for comparison.

To ensure all queries complete in a reasonable time, we set timeouts: 3 hours for small queries and 30 minutes for large queries. Queries exceeding the limit are marked as unsolved, and we report both the number of unsolved queries and the average search time. We also measure speedup using the following formula:
\[
\text{Speedup}(\mathcal{A} \text{ over } \mathcal{B}) = \frac{1}{|\mathcal{Q}|} \sum\nolimits_{Q \in \mathcal{Q}} \frac{t_{\mathcal{B}}(Q)}{t_{\mathcal{A}}(Q)},
\]
where $\mathcal{Q}$ is the set of queries, and $t_{\mathcal{X}}(Q)$ is the search time of algorithm $\mathcal{X}$ on query $Q$. We also evaluate GPU memory consumption. Since $G$ occupy the same amount of memory across all methods, we focus on measuring the space consumed by the execution stack.

\subsection{Overall Comparison}
\label{sec:overall_comparison}

\textbf{Small Query.} We evaluate \textph{} against STMatch, T-DFS, BEEP and G$^2$Miner on the six large datasets. EGSM is excluded since it targets small data graphs and frequently runs out of memory on these datasets. The result is shown in Table~\ref{tab:g2}. For fairness, we disable auxiliary structures and BFS in \textph{}, comparing only DFS-based search time with the baselines. \sun{gMatch is the only method that completes all queries across all datasets. In contrast, the baselines encounter OOM or OOT failures on large graphs (e.g., \emph{fr} and \emph{rm}) or graphs with high maximum degree (e.g., \emph{ld}), demonstrating the superior scalability and generality of our approach.}

STMatch frequently runs out of memory because it allocates a buffer of size $d_{\max}$ at each stack level. In particular, it fails on \emph{ld}, where $d_{\max}$ exceeds 4 million, and on \emph{fr} and \emph{rm}, where the remaining memory after stack allocation is insufficient. T-DFS is consistently slower than \textph{} due to the overhead of its page-table-based stack management. It also fails on \emph{fr} and \emph{rm} because representing data edges as arrays of edge pairs nearly doubles the memory footprint of the data graph, exceeding available memory.

\sun{G$^2$Miner is a graph pattern mining system that supports different tasks such as motif discovery and frequent subgraph mining, with subgraph matching as a core operation. Given a query, it employs the AutoMine compiler~\cite{10.1145/3341301.3359633} to optimize execution. However, the open-source release does not include the compiler and supports only a limited set of query patterns. As a result, our evaluation is restricted to a subset of the patterns in Figure~\ref{patterns} (P1–P4 and P9). G$^2$Miner performs well on small graphs (e.g., \emph{pk}, \emph{lj}, and \emph{ot}), but encounters OOM and OOT failures on large graphs.}

\sun{BEEP supports patterns with a single “central” vertex connected to all others (e.g., P1–P3 and P6–P9 in Figure~8) because it accelerates set intersection by constructing an adjacency matrix of size $O(d_{\max}^2)$ for the neighborhood of the matched central vertex. This matrix-based optimization enables good performance on P6–P9. However, it does not scale to large graphs and fails on \emph{ld}, \emph{fr}, and \emph{rm}. Although the source code provides an option to disable this optimization, doing so leads to incorrect results. In summary, gMatch demonstrates superior scalability and broader applicability than existing methods.}

\noindent\textbf{Large Query.} To evaluate performance on larger query graphs, we conduct our experiments on \textit{en}, \textit{gh}, \textit{gw}, \textit{db}, and \textit{wt}, as they are used in EGSM's experiments. \sun{We compare only against EGSM, since the open-source implementations of the other methods do not support arbitrary queries with tens of vertices.} For each dataset, we generate a query set consisting of 100 random 12-vertex query graphs using the method described in Section~\ref{sec:experimental_setup}. Figure~\ref{fig:egsm_comparison} demonstrates the result, where each red dot denotes a test case. Red dot below the diagonal represents one test case where \textph{} is faster than EGSM. Experiments result shows that \textph{} outperforms EGSM on all datasets, achieving 36.58$\times$ average speedup on \textit{gw} and 20.23$\times$ average speedup on \textit{db}. This speedup mainly comes from improved warp utilization enabled by the fine-grained parallel execution and warp-level batch exploration.

\noindent\textbf{Memory Consumption.} We compare gMatch’s execution-stack memory consumption with that of STMatch and T-DFS. Figure~\ref{fig:memory_consumption} shows the maximum memory consumption of three algorithms on each dataset, where each OOM denotes an out-of-memory error. We can see that STMatch incurs severe memory overhead. While T-DFS effectively reduces memory cost using a page table, it still consumes more memory than \textph{}. \textph{} minimizes the stack size to fit within GPU shared memory, requiring no additional global memory for the execution stack.

\begin{table}[htbp]
  \centering
  \small
  \caption{\textcolor{black}{Search time (millisecond) for all patterns across datasets. Each value represents the average search time over the \emph{lj}, \emph{ot}, \emph{pk}, and \emph{fr} datasets.}}
  \label{tab:avg_ablation}
  \begin{tabular}{lccccc}
    \toprule
    & \makecell[c]{\textbf{Naive}} & \makecell[c]{\textbf{Unroll-4}} & \makecell[c]{\textbf{Unroll-8}} & \makecell[c]{\textbf{Fine}\\\textbf{(naive)}} & \makecell[c]{\textbf{Fine}\\\textbf{(batch)}} \\
    \midrule
    P1 & 1,355 & 1,277 & 1,221 & 1,249 & \textbf{1,132} \\
    P2 & 4,334 & 4,098 & 3,917 & 4,058 & \textbf{3,429} \\
    P3 & 20,284 & 19,022 & 18,443 & 18,855 & \textbf{15,905} \\
    P4 & 935,433 & 877,175 & 849,963 & 869,191 & \textbf{714,827} \\
    P5 & 617,268 & 578,856 & 560,772 & 573,618 & \textbf{483,227} \\
    P6 & 924,871 & 866,658 & 839,577 & 858,787 & \textbf{743,052} \\
    P7 & 418,832 & 392,459 & 380,196 & 388,898 & \textbf{340,906} \\
    P8 & 186,048 & 174,420 & 168,945 & 174,327 & \textbf{140,781} \\
    P9 & 21,460 & 20,106 & 19,487 & 19,922 & \textbf{16,785} \\
    \bottomrule
  \end{tabular}
\end{table}

\subsection{Evaluation of Individual Techniques} \label{sec:individual_evaluation}

We evaluate the effectiveness of fine-grained parallel execution and warp-level batch exploration. We compare the performance of all query graphs from Figure~\ref{patterns}. There are five methods under study: (1) Naive coarse-grained model; (2) Coarse-grained model with loop unrolling size of 4; (3) Coarse-grained model with loop unrolling size of 8; (4) Fine-grained model without batch exploration; and (5) Fine-grained model with batch exploration. We evaluate the performance of the five methods across six datasets: \textit{lj}, \textit{ot}, \textit{ld}, \textit{pk}, \textit{fr}, and \textit{rm}. The average search time for each query is reported in Table~\ref{tab:avg_ablation}. Note that \textit{ld} and \textit{rm} are excluded from the average time calculation because the Naive, Unroll-4, and Unroll-8 methods fail on these datasets.

The experimental results show that the fine-grained model with batch exploration consistently achieves the best performance. Only our fine-grained methods successfully handle the \textit{ld} and \textit{rm} datasets, which have high maximum degrees. All coarse-grained approaches fail on these challenging datasets due to out-of-memory errors.

\begin{figure}[t]
\centering
\begin{minipage}{0.49\linewidth}
    \centering
    \includegraphics[width=\linewidth]{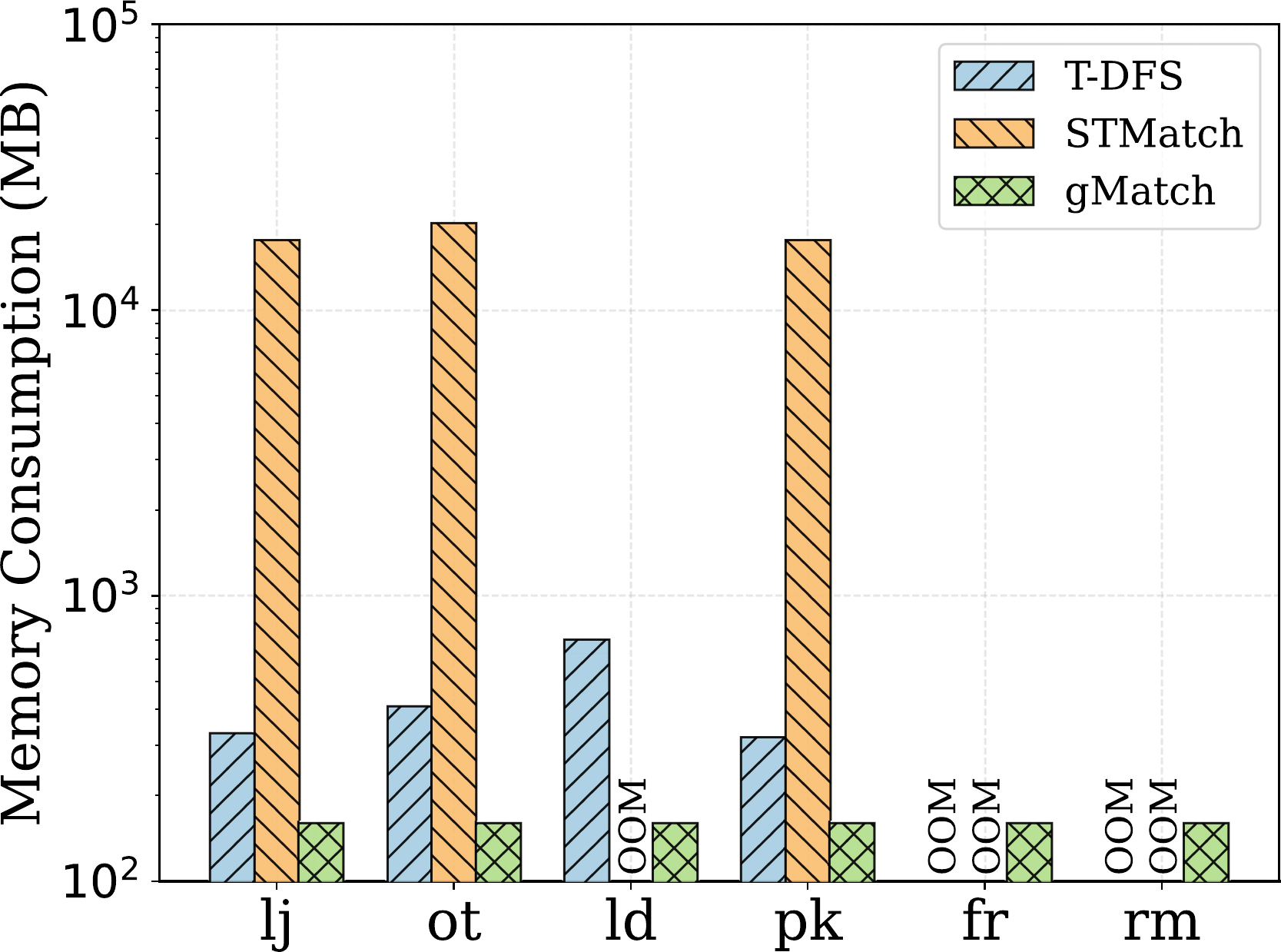}
    \caption{\textcolor{black}{Memory consumption of the execution stacks during DFS.}}
    \label{fig:memory_consumption}
\end{minipage}
\hfill
\begin{minipage}{0.49\linewidth}
    \centering
    \includegraphics[width=\linewidth]{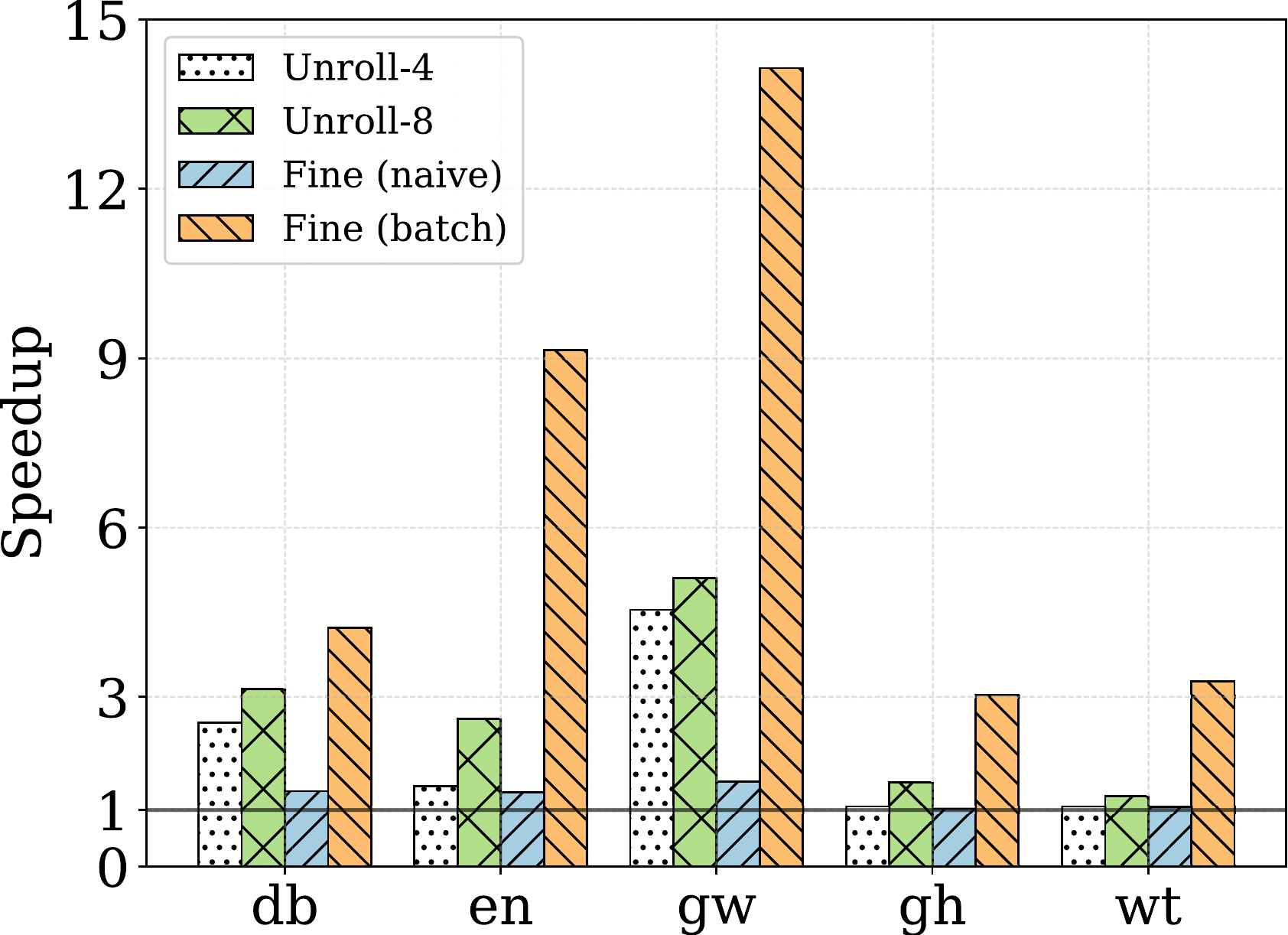}
    \caption{Speedup comparison vs. naive coarse-grained parallel execution.}
    \label{fig:unroll_time}
\end{minipage}
\end{figure}

To evaluate the effectiveness of fine-grained parallel execution on large queries over medium-sized data graphs, we measure the average speedup on random query sets of 12 vertices (Figure~\ref{fig:unroll_time}). The baseline is the coarse-grained parallel approach without loop unrolling. Our batch exploration strategy achieves significant speedup because candidate sets are  small after filtering, causing substantial thread idling in coarse-grained models.

To quantify this effect, we measure the idle rate (defined in Section~\ref{sec:issues}) across the datasets, as shown in Table~\ref{tab:my_idle_rate}. Compared with existing methods, our approach reduces the idle rate to below 5\%, demonstrating much higher efficiency.

\begin{table}[t]
  \centering
  \small
  \caption{Comparison of idle rates between \textph{} and the baseline STMatch under its different loop unrolling configurations (naive, unroll-2, unroll-4, unroll-8).}
  \label{tab:my_idle_rate}
  \begin{tabular}{cccccc}
    \toprule
    & \textbf{\textit{db}} & \textbf{\textit{en}} & \textbf{\textit{gw}} & \textbf{\textit{gh}} & \textbf{\textit{wt}} \\
    \midrule
Naive         & 70.74\% & 45.14\% & 40.34\% & 48.19\% & 58.80\% \\
Unroll-2         & 50.22\% & 31.38\% & 38.62\% & 59.47\% & 29.82\% \\
Unroll-4         & 30.16\% & 20.75\% & 22.94\% & 21.62\% & 12.49\% \\
Unroll-8         & 19.32\% & 10.84\% & 15.54\% & 11.41\% &  9.80\% \\
\textph{} & 4.14\% & 3.41\% & 3.53\% & 4.04\% & 1.51\% \\
    \bottomrule
  \end{tabular}
\end{table}

\begin{figure}[b]
\begin{subfigure}[b]{0.48\linewidth}
\includegraphics[width=\linewidth]{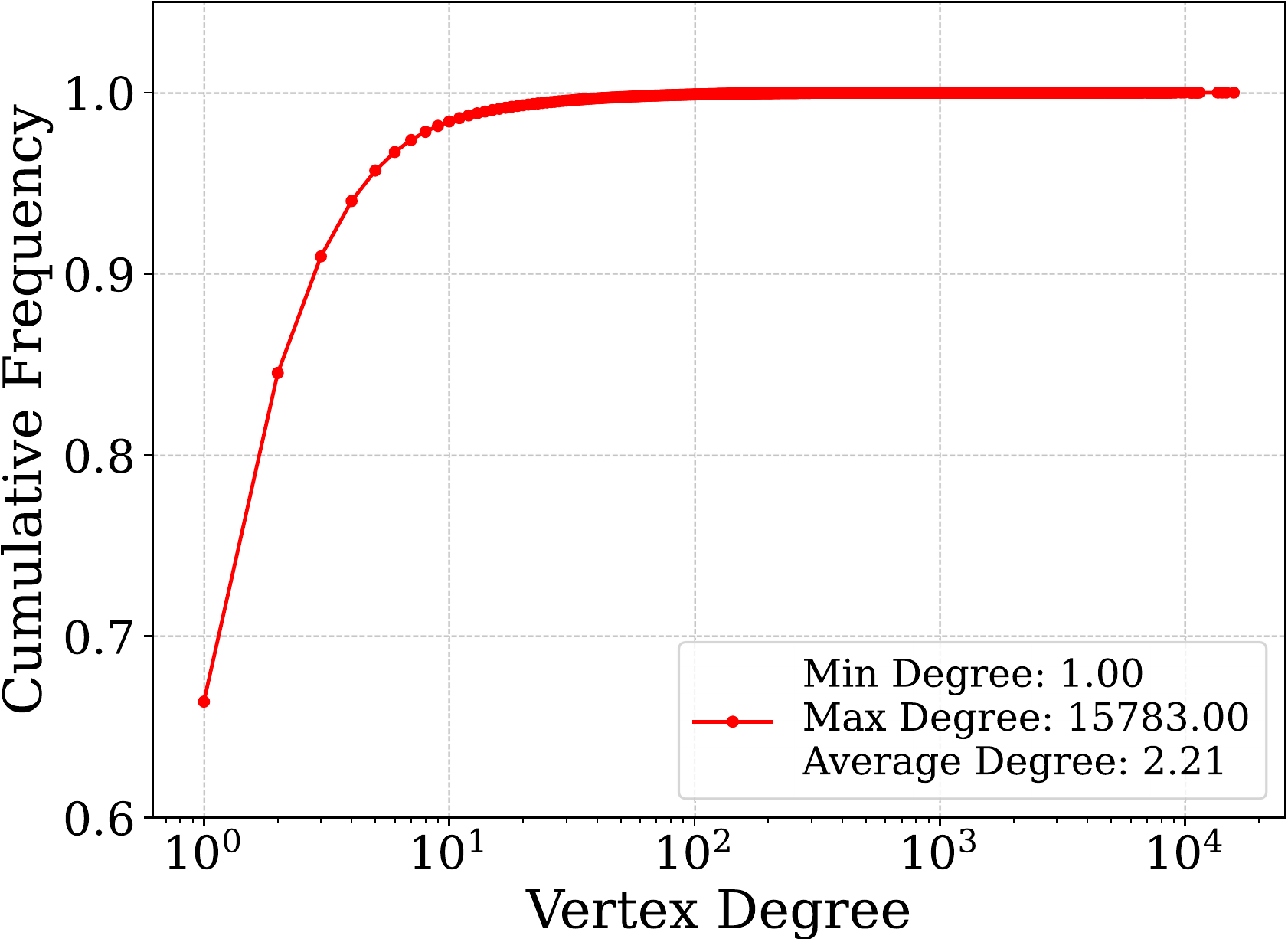}
\caption{Cumulative distribution of vertex degrees.}
\label{fig:risk_degree}
\end{subfigure}
\hfill
\begin{subfigure}[b]{0.49\linewidth}
\includegraphics[width=\linewidth]{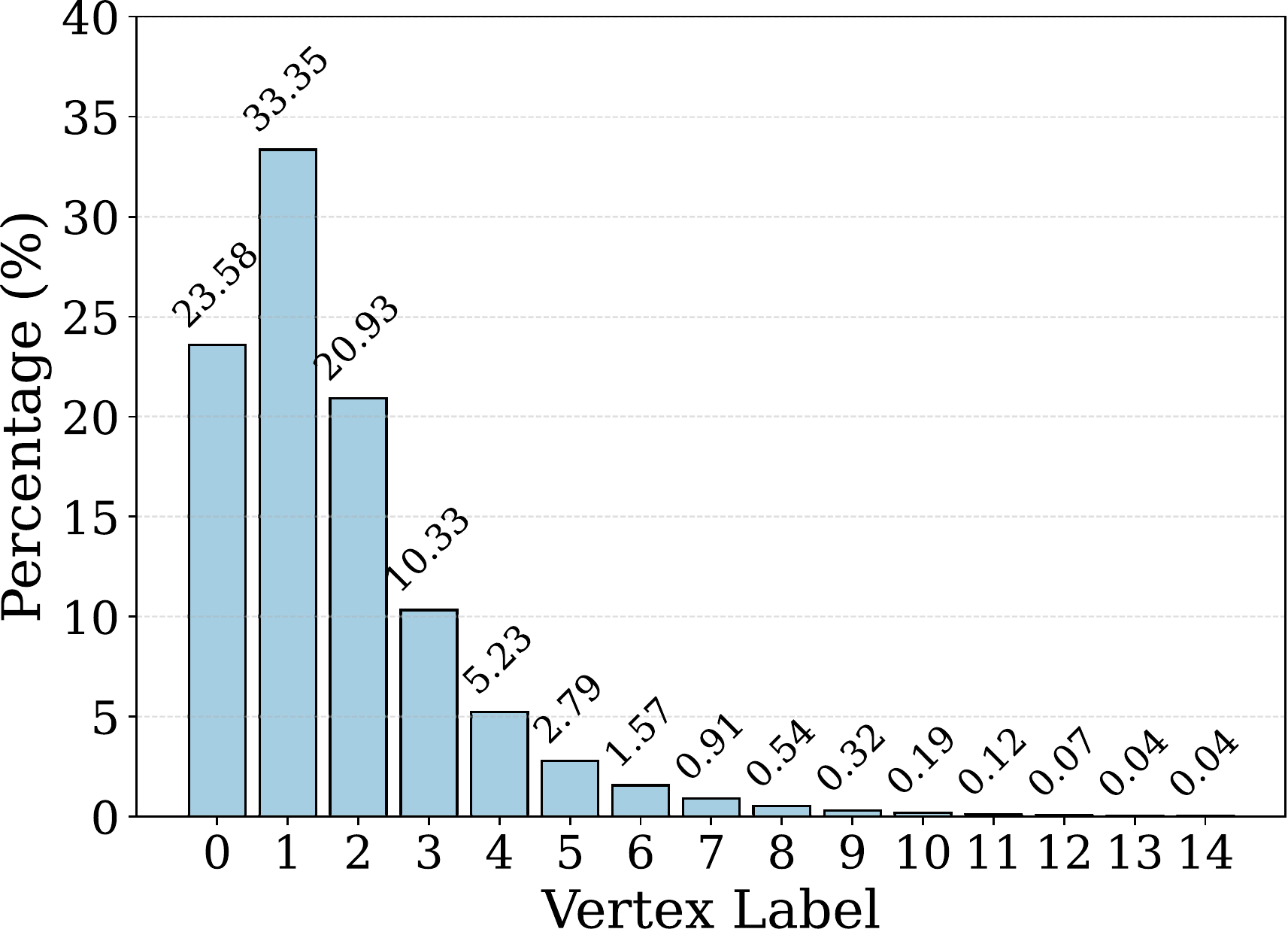}
\caption{Distribution of vertex risk labels.}
\label{fig:risk_labels}
\end{subfigure}
\caption{Dataset of user payment relationships.}
\end{figure}

\subsection{Case Study}

We conduct a case study on a production graph representing user payment relationships from our industry partner ByteDance, one of the largest social network companies in the world. In this graph, each vertex corresponds to a user, and each undirected edge denotes a payment transaction, reflecting mutual interaction. The graph contains 139,297,601 vertices and 153,853,965 edges. Figure \ref{fig:risk_degree} shows its degree distribution. Note that different from our previous experiments, we conduct all our case study on a V100 GPU with 32 GB device memory.

\begin{figure}[htbp]
\begin{subfigure}[b]{0.475\linewidth}
\includegraphics[width=\linewidth]{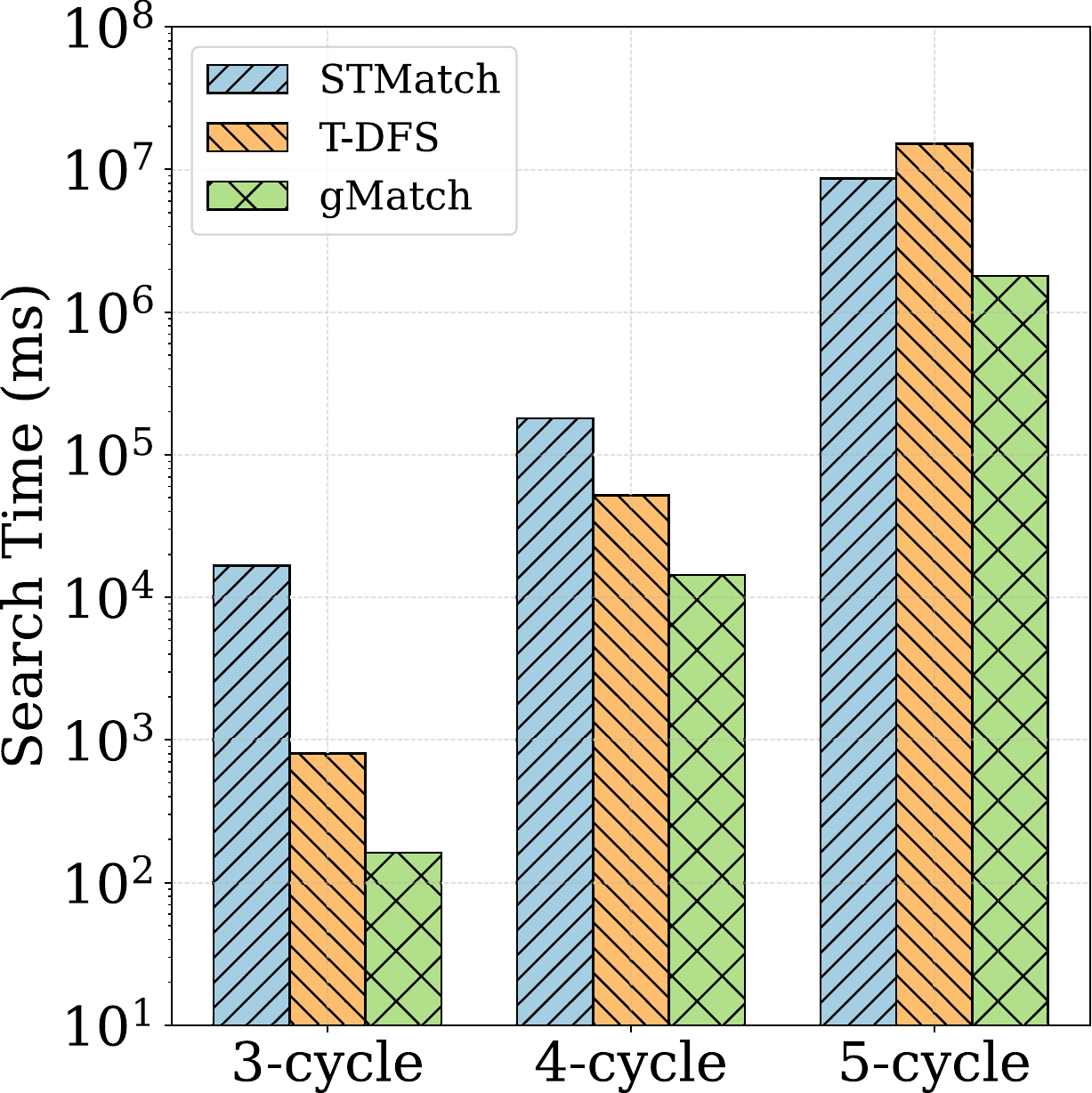}
\caption{Small queries.}
\label{fig:risk_query_time_comparison}
\end{subfigure}
\hfill
\begin{subfigure}[b]{0.49\linewidth}
\includegraphics[width=\linewidth]{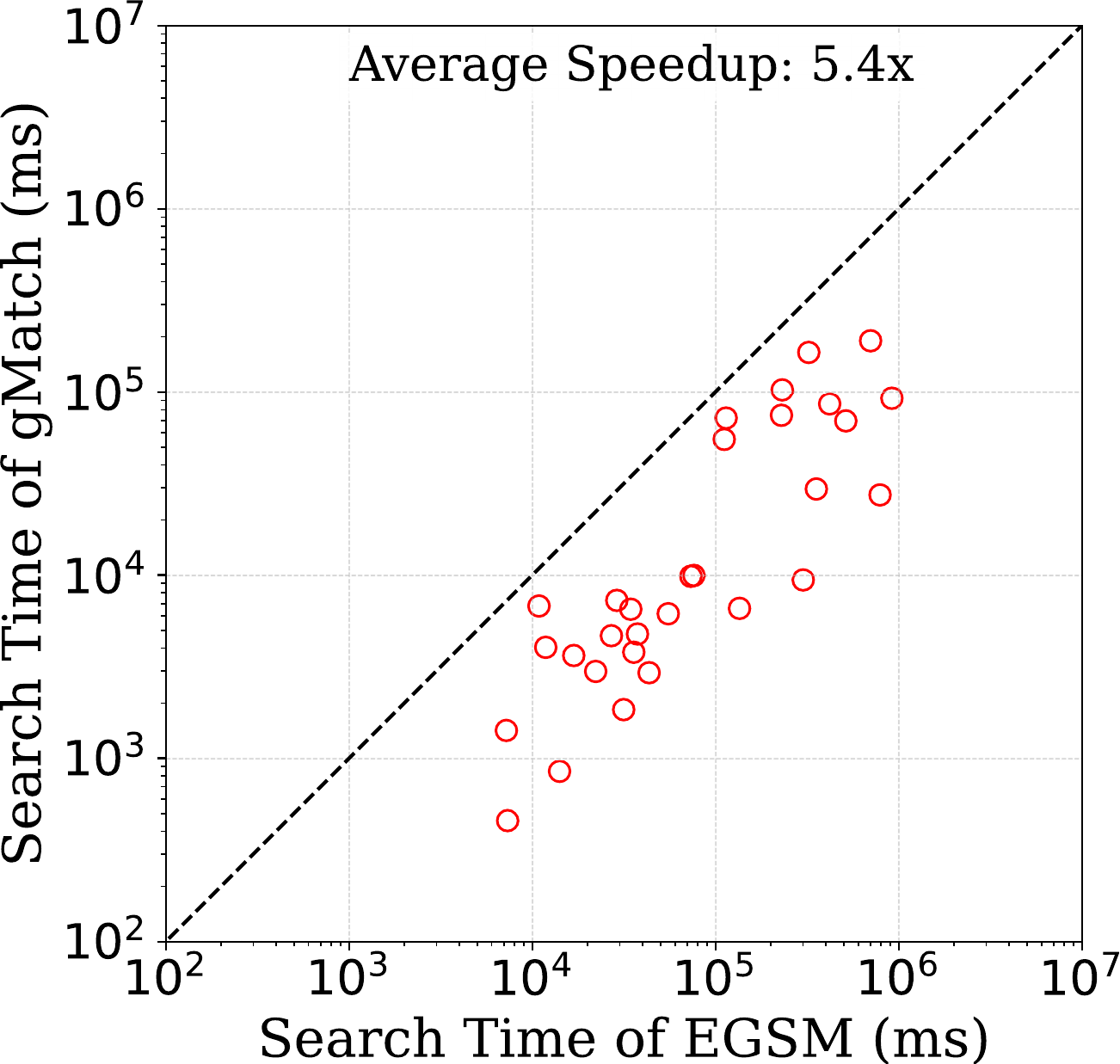}
\caption{Large queries.}
\label{fig:risk_query_egsm}
\end{subfigure}
\caption{Comparison of search times for the methods under study on the user payment network.}
\label{fig:case_study}
\end{figure}

For small queries, we focus on identifying cycles of length 3 to 5, which represent circular transaction flows. These patterns are of particular interest as they often indicate suspicious or closed-loop behaviors. The detected cycles are used as input to downstream tasks such as graph neural network (GNN) analysis. The corresponding results are shown in Figure \ref{fig:risk_query_time_comparison}.

The graph also includes vertex labels indicating user risk levels, ranging from 0 to 14. Figure \ref{fig:risk_labels} illustrates the label distribution. For large-query evaluation, we analyze transaction patterns across different risk categories using queries with dozens of vertices. Due to privacy constraints, we cannot publish the actual queries and instead generate 30 synthetic queries with 12 vertices. We generate these queries by adapting the methodology from Section \ref{sec:experimental_setup}, with one modification: to ensure the inclusion of risky vertices, the seed vertex is always randomly selected from the set of vertices whose labels fall within the range of 10 to 14. Figure \ref{fig:risk_query_egsm} presents the results of our case study. As shown in the figure, our method significantly outperforms the existing approaches.

\section{Related Work} \label{sec:related_work}

\noindent\textbf{GPU-based subgraph matching.} Recent research on subgraph matching has increasingly leveraged GPUs due to their massive parallelism capabilities. Existing GPU-based subgraph matching solutions can be categorized into BFS-based, DFS-based, and hybrid approaches according to their search strategy. Early GPU subgraph matching research primarily adopted BFS-based methods due to their inherent workload balance advantages \cite{gpsm,gsi,cuts}. However, BFS-based approaches typically generate a large amount of intermediate results that cannot fit in the memory of GPUs. To reduce memory consumption, recent studies have revisited DFS-based strategies \cite{tdfs,stmatch,parsec,BEEP} or DFS-BFS hybrid strategies \cite{egsm}. Moreover, there are some researches focusing on processing subgraph matching on very large data graphs that cannot fit in GPU memory. They either divide the data graph into multiple partitions \cite{pbe} or use a view-based strategy according to the matched data vertex of the source query vertex \cite{vsgm,SGSI,g2aimd}. These algorithms preprocess the data graph by dividing the problem into multiple chunks and load one chunk into memory at a time. Additionally, there are GPU algorithms specifically designed for $k$-clique counting \cite{k_clique_gpu,clique-2} and triangle counting \cite{triangle-1,triangle-2,triangle-3}, which are special cases of subgraph matching. While these works demonstrate high performance, they are not applicable to matching general patterns.

\noindent\textbf{CPU-based subgraph matching.} Existing CPU-based subgraph matching algorithms primarily leverage pruning strategies and auxiliary data structures to enhance performance. The solutions of subgraph matching can be roughly divided into join-based and exploration-based, as indicated by \cite{study}. Current exploration-based approaches are fundamentally rooted in Ullmann's algorithms \cite{ullmann}, incrementally matching query vertices to candidate vertices. Subsequent researches effectively reduce the search space by using optimized matching orders \cite{ri,GraphQL}, applying advanced filtering techniques \cite{rapidmatch,versatile,fastcsm,ceci}, avoid redundant computations \cite{circinus,BSX}, and maintaining failure sets \cite{Harmonizing,gup}. While these exploration-based methods excel at handling complex queries on single machines through their complicated pruning strategies, join-based approaches offer better scalability for large-scale datasets by decomposing the query into sub-structures that can be processed in parallel, making them better suited for distributed environments.

\noindent\textbf{Graph pattern mining.} Graph pattern mining (GPM) focuses on finding specific patterns within large data graphs. It includes several applications such as subgraph listing \cite{10.1145/2882903.2915209,g2miner}, k-motif counting \cite{10.14778/3364324.3364330,g2miner}, and k-frequent subgraph mining \cite{10.14778/3389133.3389137,g2miner}. Early work tackled these problems by creating specialized, custom solutions for each specific task. This strategy is effective but not general-purpose. To make it easier to run different mining tasks efficiently in parallel, general-purpose graph mining systems were developed. These systems used an embedding-centric approach, starting with small pieces and gradually extending them step-by-step into larger subgraphs, checking at each stage if they meet the user's criteria. Systems like \cite{10.1145/2815400.2815410,10.1145/3190508.3190545} used this model. However, similar to subgraph matching algorithms, these GPM systems also face the problem of tracking the huge number of intermediate partial results. To solve this problem, \cite{10.1145/3299869.3319875,10.1145/3341301.3359633,g2miner} adopt DFS strategies for enumeration.

\section{Conclusion} \label{sec:conclusion}

We presented \textph{}, a fine-grained and hardware-efficient subgraph matching method for GPUs. Our work is driven by the key insight that the coarse-grained
parallel execution model fundamentally conflicts with real-world
power-law graphs, giving rise to two inherent issues. First, GPU
scalability is constrained by memory: large per-task stacks severely limit the number of concurrent warps and
underutilize GPU parallelism, making a lightweight stack design
essential. Second, mapping highly irregular graph workloads onto
fixed-size warps leads to significant thread underutilization. To
address these challenges, our fine-grained design incorporates
warp-level batch exploration and a lightweight load-balancing mechanism,
achieving high efficiency and scalability across a wide range of
workloads. Experimental results show that \textph{} delivers strong performance and scalability with low memory overhead, making it well suited for practical, large-scale graph applications.

\begin{acks}
The work is supported by the National Natural Science Foundation of China (NSFC) under Grant No. 62572303. Shixuan Sun and Minyi Guo are the corresponding authors.
\end{acks}

\newpage

\balance
\bibliographystyle{ACM-Reference-Format}
\bibliography{ref}
\balance

\newpage

\appendix

\nobalance
\FloatBarrier

\section{Scalability Evaluations} \label{sec:appendix_a}

To evaluate the scalability of our algorithm, we conduct four experiments on two representative datasets (\textit{db} and \textit{gw}) and one experiment using the LDBC benchmark. We consider five aspects: query graph density, query graph size, label cardinality, skewness of data vertex label, and data graph size. For analysis of data graph sizes, we conduct this experiment exclusively with T-DFS, as EGSM fails to process large data graph and STMatch does not support large maximum degree. We conduct the rest of scalability evaluations exclusively with EGSM, since the released code of STMatch and T-DFS only support small query graphs with no more than 7 vertices.

\begin{figure}[htbp]
    \centering
    \begin{subfigure}[b]{0.49\linewidth}
        \centering
        \includegraphics[width=\linewidth]{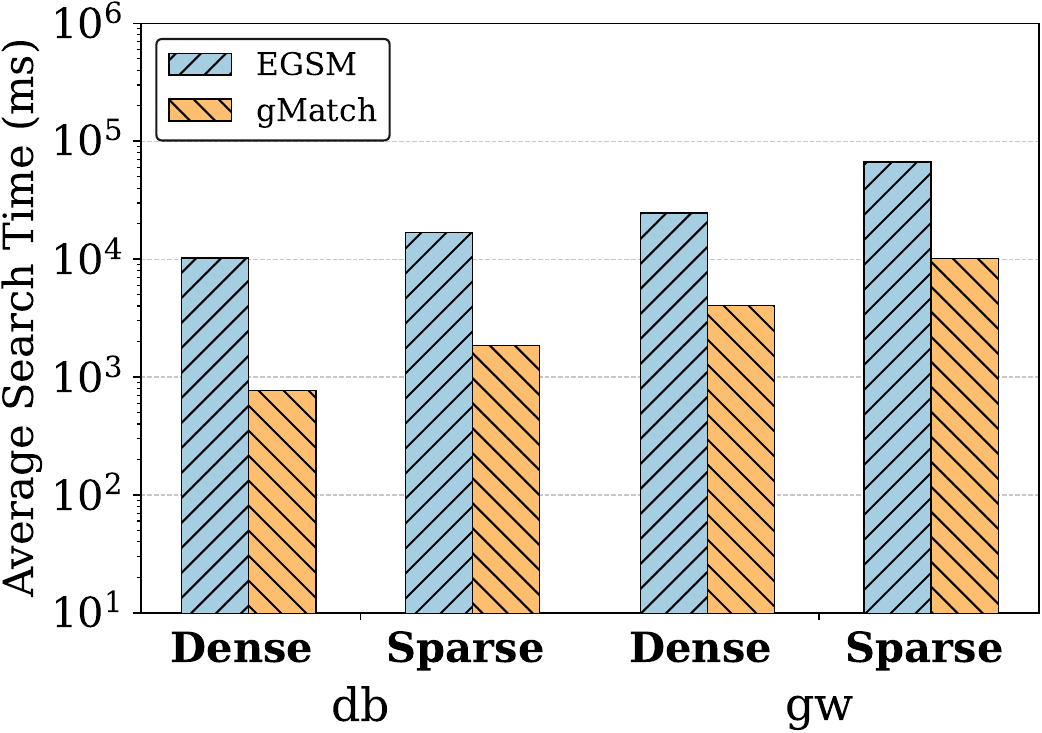}
        \caption{Search time with varying $Q$ density.}
        \label{fig:aaa}
    \end{subfigure}
    \hfill
    \begin{subfigure}[b]{0.49\linewidth}
        \centering
        \includegraphics[width=\linewidth]{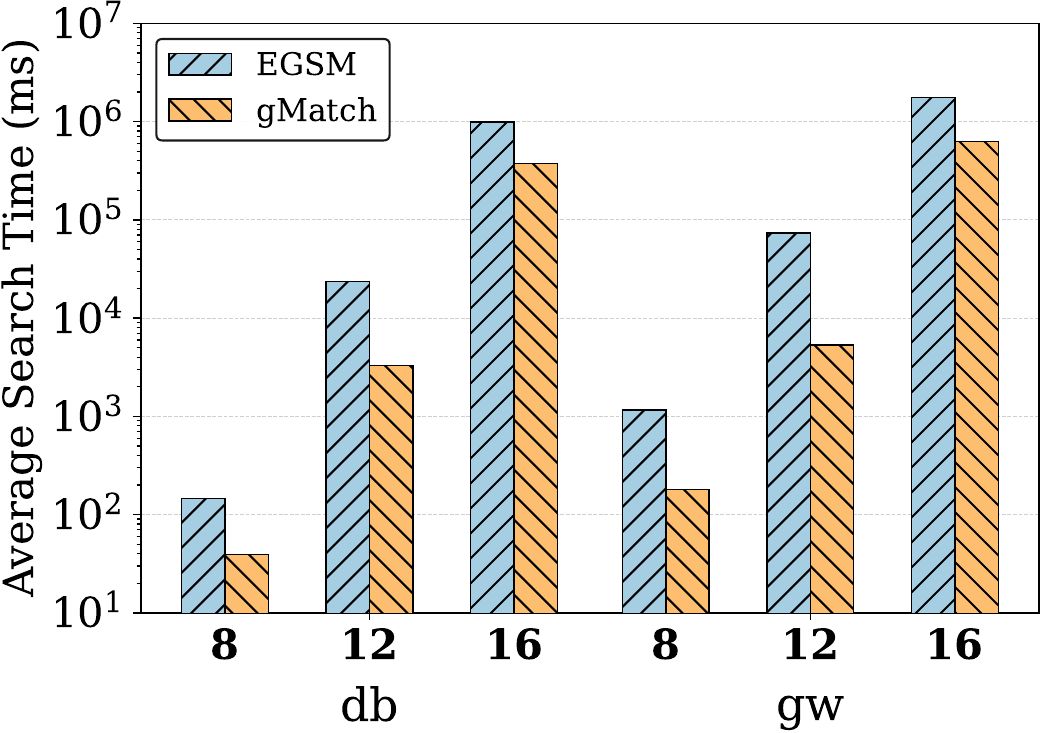}
        \caption{Search time with varying $|V(Q)|$.}
        \label{fig:bbb}
    \end{subfigure}
    
    \vspace{0.5cm}
    
    \begin{subfigure}[b]{0.49\linewidth}
        \centering
        \includegraphics[width=\linewidth]{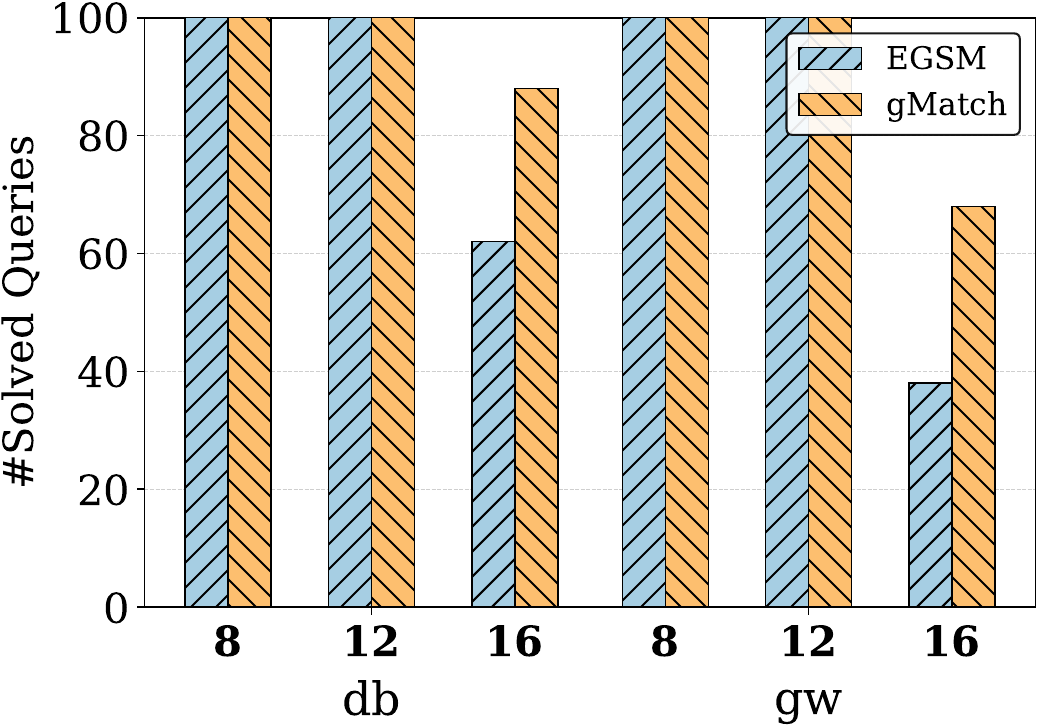}
        \caption{Number of solved queries with varying $|V(Q)|$.}
        \label{fig:ccc}
    \end{subfigure}
    \hfill
    \begin{subfigure}[b]{0.49\linewidth}
        \centering
        \includegraphics[width=\linewidth]{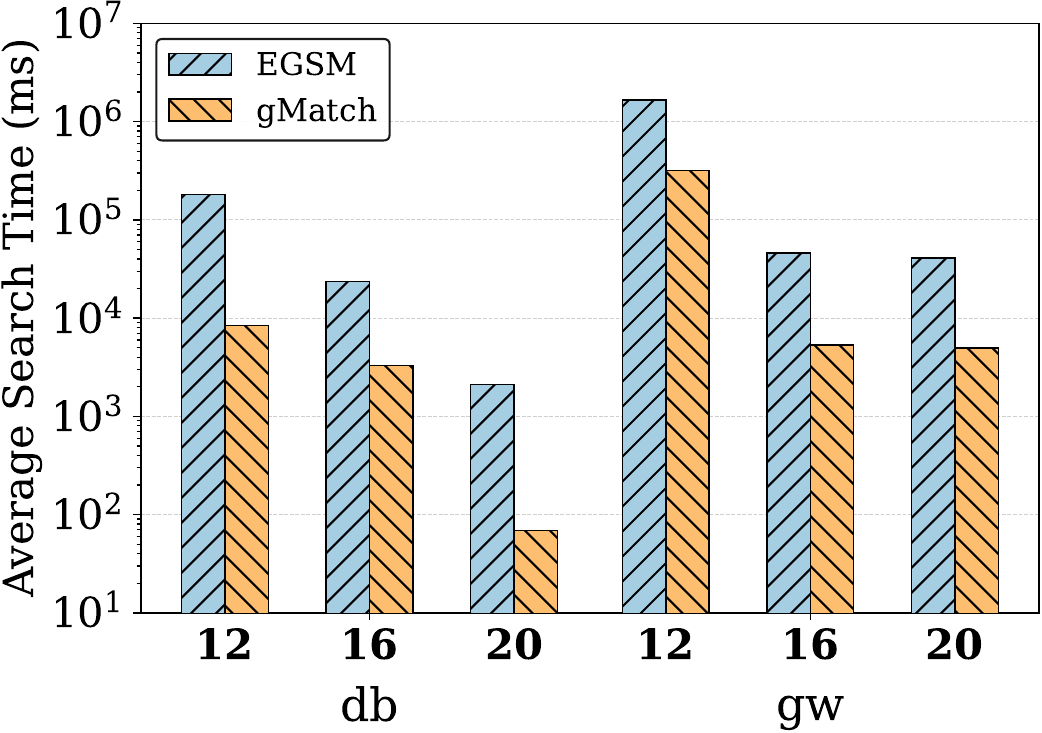}
        \caption{Search time with varying $|\Sigma(G)|$.}
        \label{fig:ddd}
    \end{subfigure}
    
    \vspace{0.5cm}
    
    \begin{subfigure}[b]{0.49\linewidth}
        \centering
        \includegraphics[width=\linewidth]{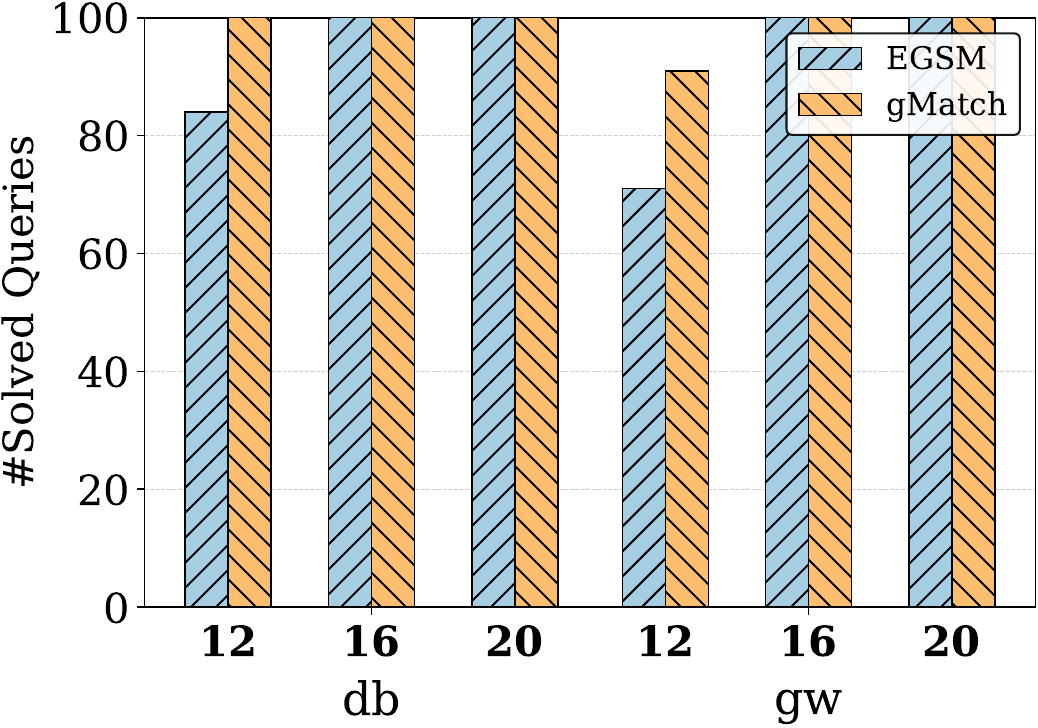}
        \caption{Number of solved queries with varying $|\Sigma(Q)|$.}
        \label{fig:eee}
    \end{subfigure}
    \hfill
    \begin{subfigure}[b]{0.49\linewidth}
        \centering
        \includegraphics[width=\linewidth]{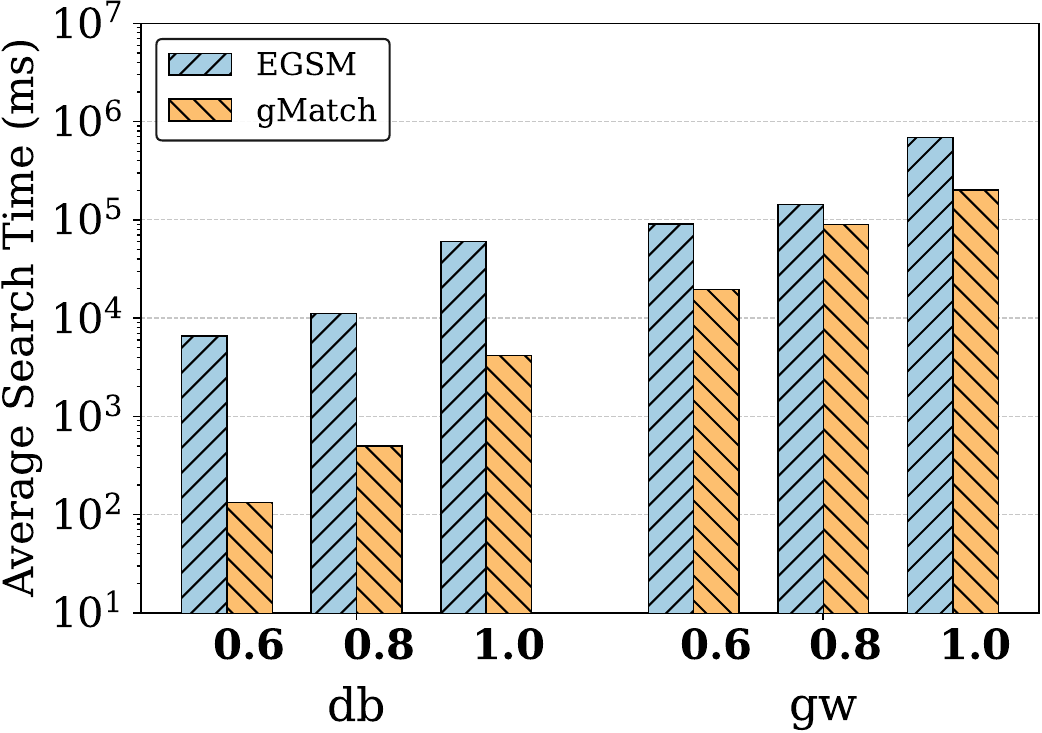}
        \caption{Search time with varying data graph skewness.}
        \label{fig:fff}
    \end{subfigure}
    
    \caption{Scalability evaluations.}
    \label{fig:scalability}
\end{figure}

\noindent\textbf{Analysis of Query Graph Density.} We first examine how query graph density impacts performance. Using 100 query graphs with 12 vertices per dataset, we categorize query graphs based on average degree (\(d_\text{avg}\)): sparse (\(d_\text{avg} < 3\)) and dense (\(d_\text{avg} \geq 3\)). As shown in Figure \ref{fig:aaa}, the query time is generally longer with smaller query graph density. This is because sparse query graphs usually lead to larger intermediate results and are more difficult to solve.

\noindent\textbf{Analysis of Query Graph Size.} To evaluate the impact of query graph size, we generate 100 random query graphs per dataset at each of three sizes: 8, 12 and 16 vertices. The average query time of solved queries is shown in Figure \ref{fig:bbb}, and the number of solved queries is shown in Figure \ref{fig:ccc}. Note that all queries with size 8 or 12 can be solved within the time limit. Typically, a larger query graph is more difficult to solve than smaller ones.

\noindent\textbf{Analysis of Label Cardinality.} To examine how the number of distinct labels \(|\Sigma(G)|\) affects performance, we use the configurations with \(|\Sigma(G)| = 12, 16,\) and \(20\). For each configuration, we generate 100 random 12-vertex queries. The average query time of solved queries is shown in Figure \ref{fig:ddd}, and Figure \ref{fig:eee} shows the number of solved queries. Note that all queries under $|\Sigma(G)|=16,20$ are solved. Generally speaking, a larger label cardinality leads to simpler subgraph matching, because this leads to smaller candidates set and thus increases the selectivity.

\noindent\textbf{Analysis of Label Distribution Skewness.} We evaluate how skewed label distributions impact performance by synthetically modifying labels of \textit{db} and \textit{gw} using Zipf's law. For each dataset, we first fix the label set size \(|\Sigma(G)|=16\), then assign vertex labels according to this distribution: $P(\text{label} = \ell) = \frac{(\ell+1)^{-\alpha}}{\sum_{i=1}^{|\Sigma|} i^{-\alpha}}$, where \(\ell\) is the label (an integer within [0,\(|\Sigma|-1\)]) and \(\alpha\) controls skewness. We test \(\alpha = \{0.6, 0.8, 1.0\}\), generating progressively skewed distributions. Specifically, a larger $\alpha$ denotes more skewed distribution. To ensure that all queries can be solved within the time limit, we execute 100 random query graphs with 10 vertices for each configuration. The result is shown in Figure \ref{fig:fff}.

\noindent\textcolor{black}{\textbf{Analysis of Data Graph Sizes.} To evaluate how gMatch and competing methods perform as data graph size increases, we use the LDBC benchmark to generate unlabeled, undirected data graphs of varying sizes, using LDBC's standard scale factors of 3, 10, 30, and 100. These parameters directly determine the graph size and were chosen to ensure that all graphs fit within the memory of a single GPU, as all evaluated methods operate in an in-memory setting. The key statistics of the generated graphs are summarized in Table \ref{tab:ldbc_datasets}}.

\textcolor{black}{Given that the large $d_\text{max}$ in the LDBC dataset causes STMatch to run out of memory, we exclude it and only evaluate gMatch and T-DFS in this experiment. Similar to Section \ref{sec:overall_comparison}, we use the same matching order and symmetry breaking strategy in gMatch and T-DFS for fairness of comparison.}

\begin{figure}[H]
\centering
\includegraphics[width=\linewidth]{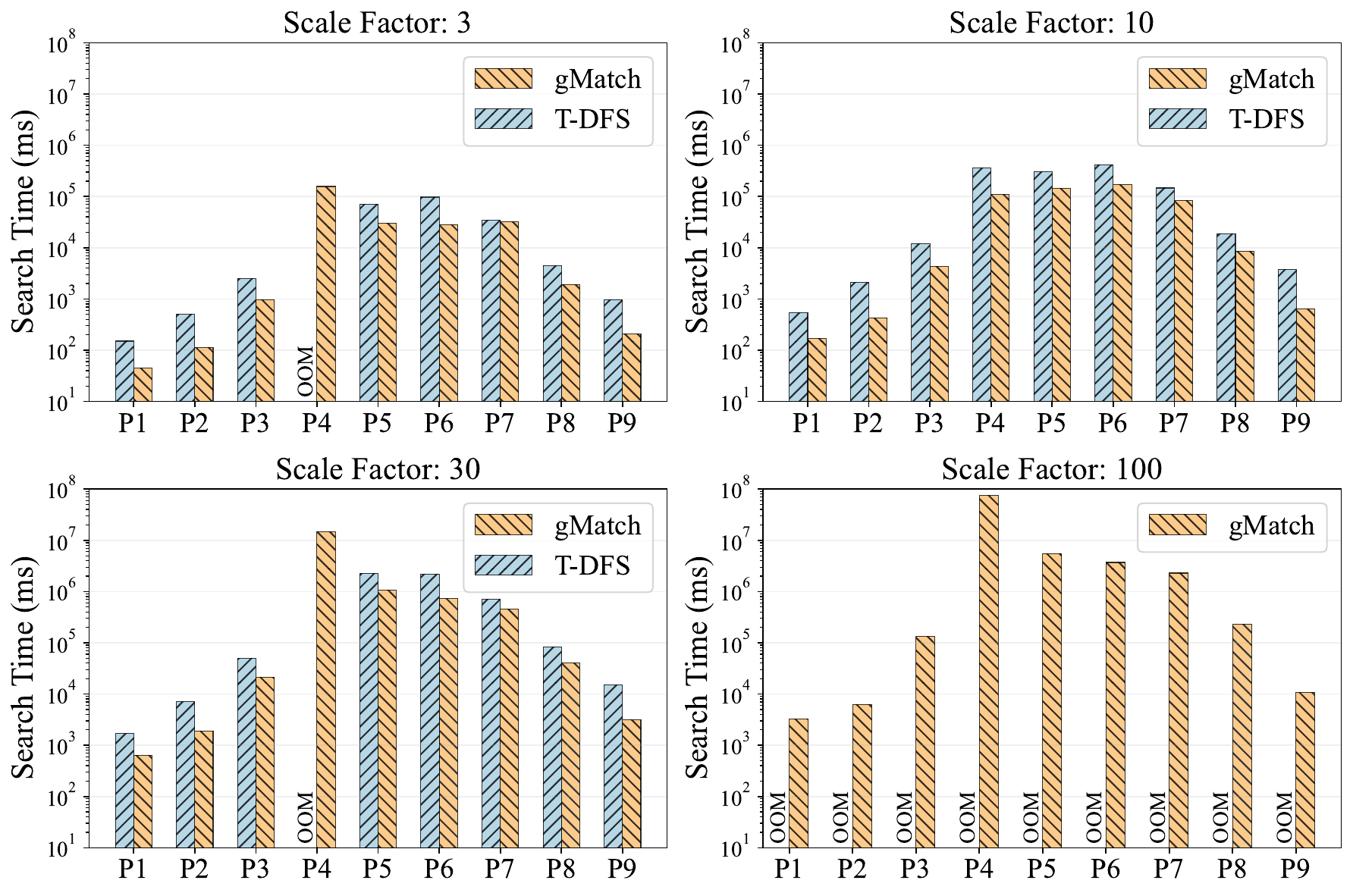}
\caption{\textcolor{black}{Search time with varying data graph sizes.}}
\label{fig:ldbc_datasets}
\end{figure}

\textcolor{black}{As shown in Figure \ref{fig:ldbc_datasets}, gMatch demonstrates robust scalability across data graphs of varying sizes. In contrast, T-DFS is slower than gMatch in all cases and, moreover, failed to complete execution for all patterns at scale factor 100 and for pattern P4 at scale factors 3 and 30. This failure is due to excessive memory allocation by its memory management system. These results collectively confirm the efficiency and scalability of gMatch.}

\begin{table}[htbp]
\caption{\textcolor{black}{The detailed statistics of LDBC datasets.}}
\label{tab:ldbc_datasets}
\centering
\resizebox{\linewidth}{!}{
\begin{tabular}{ccccc}
   \toprule
   \textbf{Scale Factor} & $|V|$ & $|E|$ & $d_\text{avg}$ & $d_\text{max}$ \\
   \midrule
   3 & 9,281,922 & 52,651,300 & 11.34 & 1,346,287 \\
   10& 29,982,730 & 175,860,387 & 11.73 & 4,282,595 \\
   30& 88,789,833 & 540,506,176 & 12.18 & 12,684,688 \\
   100 & 282,386,021 & 1,773,425,640 & 12.56 & 40,767,884 \\
   \bottomrule
\end{tabular}}
\end{table}

\section{Evaluations of Load Balancing} \label{sec:appendix_b}

We evaluate the impact of the initialization phase and work stealing in our load-balancing strategy, respectively.

\begin{figure}[htbp]
\begin{subfigure}[b]{0.49\linewidth}
\includegraphics[width=\linewidth]{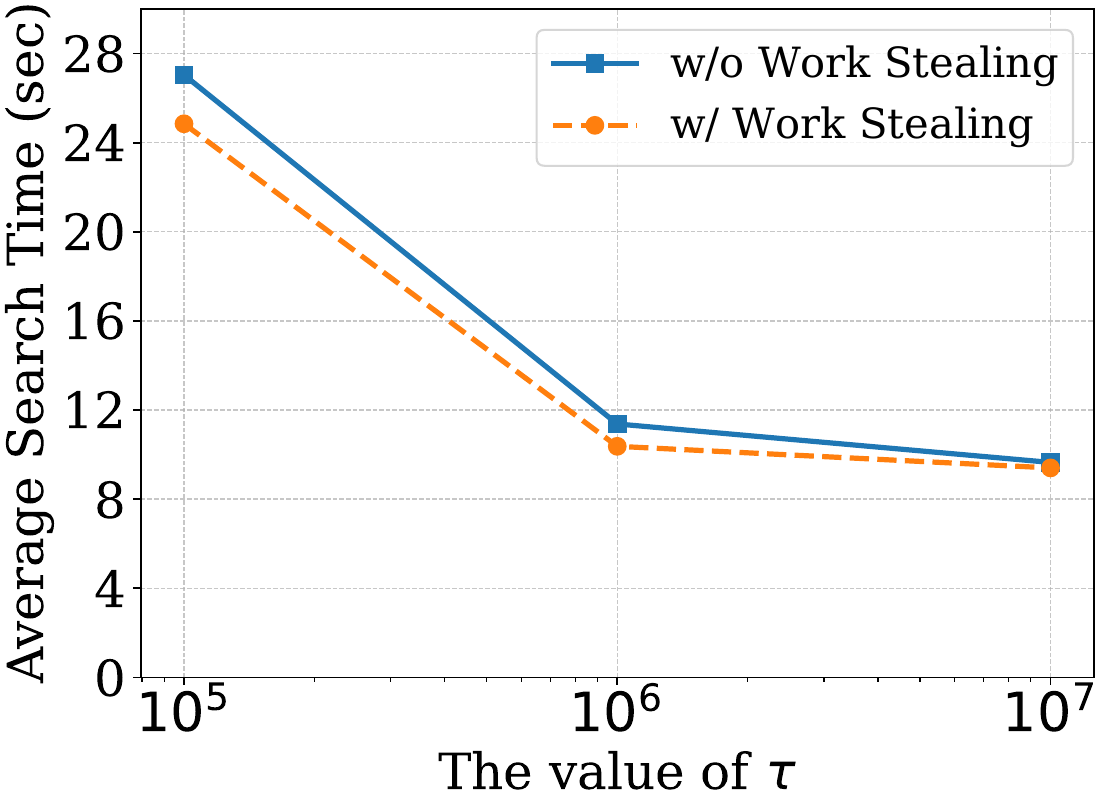}
\caption{On the dataset \textit{en}.}
\label{fig:enron_work_stealing_speedup}
\end{subfigure}
\hfill
\begin{subfigure}[b]{0.49\linewidth}
\includegraphics[width=\linewidth]{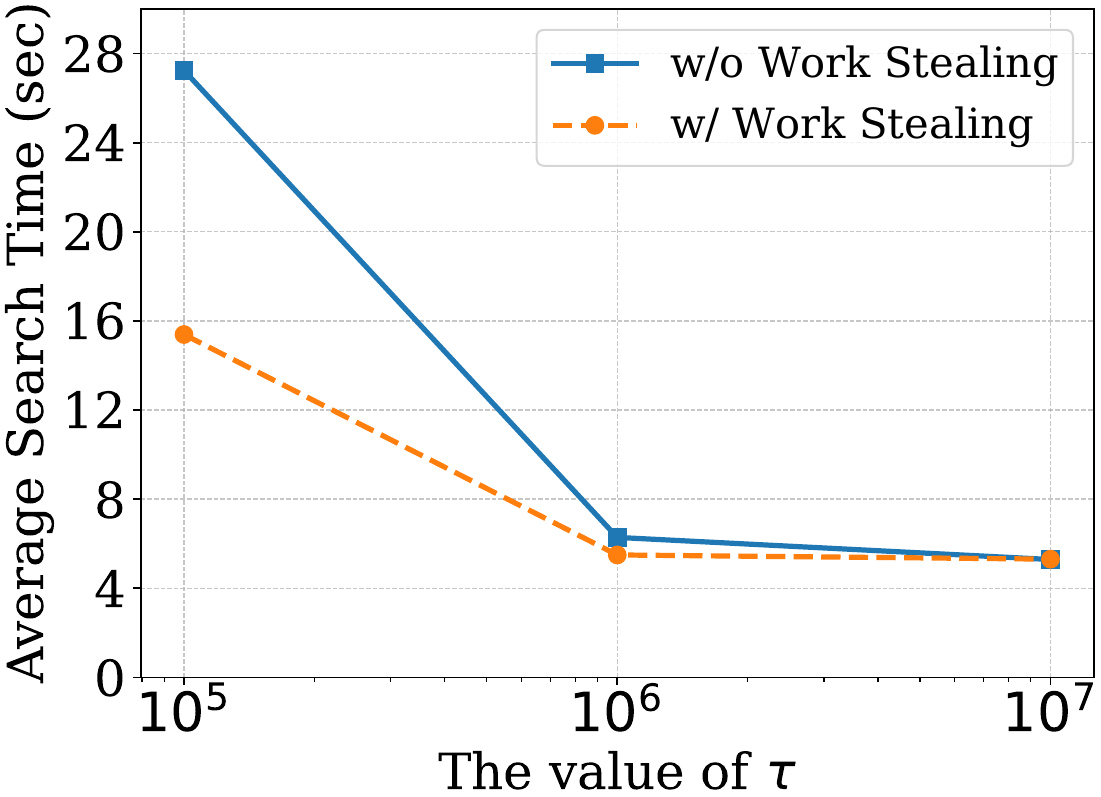}
\caption{On the dataset \textit{gw}.}
\label{fig:gowalla_work_stealing_speedup}
\end{subfigure}
\caption{Search time under different initial task sizes. Each value represents the average search time over 100 queries with 12 query vertices on the dataset.}
\label{fig:work_stealing_speedup}
\end{figure}

\emph{Evaluation of Initialization Phase.} To evaluate the impact of the initial task pool size $\tau$, we conduct experiments on the \textit{en} and \textit{gw} datasets using multiple $\tau$ values ($10^5$, $10^6$, $10^7$). Values below $10^5$ cause significant load imbalance because some warps may receive no initial tasks, so smaller thresholds are excluded. We also compare configurations with work stealing enabled and disabled to assess its effect under different $\tau$ settings. The results are shown in Figure~\ref{fig:work_stealing_speedup}. Larger $\tau$ values improve load balance by increasing the number of initial tasks, enabling dynamic task fetching to better mitigate imbalance. When $\tau$ exceeds $10^6$, the performance with and without work stealing becomes similar, indicating that the initialization phase alone provides sufficient load balance.

\begin{figure}[htbp]
\centering
\includegraphics[width=0.7\linewidth]{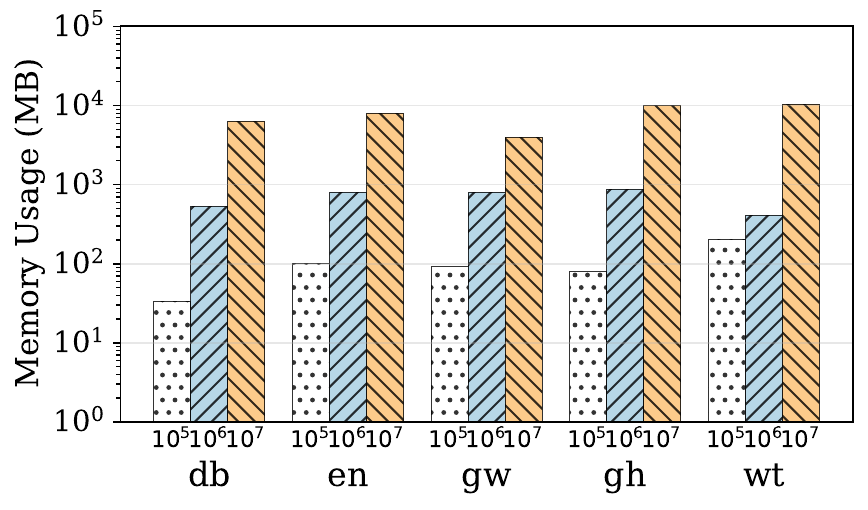}
\caption{\textcolor{black}{Maximum memory consumption of the initial task pool with varying threshold $\tau$.}}
\label{fig:memory_usage_tau}
\end{figure}

We further evaluate the memory consumption on varying $\tau$. For each dataset, we record the maximum memory consumption of initial task pools across queries. The result is shown in Figure \ref{fig:memory_usage_tau}. When $\tau = 10^6$, the initial task pool consumes a few hundred MB, which is small compared with the GPU memory, but it grows to several GB when $\tau = 10^7$. As shown in Figure~\ref{fig:work_stealing_speedup}, increasing $\tau$ from $10^6$ to $10^7$ provides only marginal improvement. Therefore, we use $\tau = 10^6$ as the default in our experiments, as it offers near-optimal performance with low memory usage. This tuning process also shows that $\tau$ can be selected easily in practice.

\emph{Evaluation of Work Stealing.} To evaluate the effect of work stealing, we compare two scenarios, work stealing enabled and disabled, using $\tau = 10^6$. Across five datasets, we measure both the average and maximum speedup obtained when work stealing is enabled, as shown in Table~\ref{tab:ws_speedup}. Our strategy is generally effective, achieving up to a $1.18\times$ average speedup on \textit{gw}. On \textit{db}, however, work stealing introduces slight overhead because the initialization phase already provides good load balance, while enabling work stealing still incurs frequent idle-queue checks.

\begin{table}[htbp]
  \centering
  \small
  \caption{Average and maximum speedup of work stealing.}
  \vspace{-8pt}
  \label{tab:ws_speedup}
  \begin{tabular}{cccccc}
    \toprule
    & \textbf{\textit{db}} & \textbf{\textit{en}} & \textbf{\textit{gw}} & \textbf{\textit{gh}} & \textbf{\textit{wt}} \\
    \midrule
    Avg & 0.97$\times$ & 1.05$\times$ & 1.18$\times$ & 1.08$\times$ & 1.03$\times$ \\
    Max & 1.01$\times$ & 2.21$\times$ & 4.72$\times$ & 2.62$\times$ & 1.67$\times$ \\
    \bottomrule
  \end{tabular}
\end{table}

\section{Extended Discussion on Workload Patterns} \label{sec:appendix_c}

Our paper, together with prior work~\cite{study,A-Comprehensive-Survey-and-Experimental,rapidmatch}, identifies two representative workload patterns.

\emph{(1) Large queries over medium-sized graphs}. These workloads involve queries with tens of vertices evaluated on graphs with millions of edges and rich label sets. They arise in complex relational analytics, where multi-step structural patterns must be identified. Because the search space grows exponentially with query size, a substantial body of CPU-based work~\cite{versatile,rapidmatch,egsm,gsi,Harmonizing,Cartesian,ceci} focuses on efficient filtering and pruning to reduce the candidate set size. These methods typically perform a lightweight filtering phase to construct a candidate graph $G’$ using label information and query structure, and then conduct search on $G’$ rather than the original graph $G$. EGSM adopts this CPU-style approach on GPUs by constructing $G’$ and enumerating on $G’$, while T-DFS and STMatch enumerate on $G$ with a very large search space. However, after pruning, the neighbor sets $N(M[u’])$ in $G’$ are often very small, leading to poor warp utilization under coarse-grained parallelism. In addition, the large query size results in a deep search tree, making the enumeration process vulnerable to significant load imbalance on GPUs.

\emph{(2) Small queries over large-scale graphs.} These workloads consist of small queries (3–5 vertices) executed on massive graphs (tens of millions to billions of edges) with sparse label sets. They represent practical scenarios such as fraud detection or network security, where simple but important structural patterns must be located at scale. This workload category has also been studied in prior subgraph matching systems \cite{tdfs,stmatch,10.14778/3389133.3389137,g2miner,BEEP,parsec}, which leverage parallelization to accelerate computation.

Because $G$ has only a few labels and the query is small, constructing a candidate graph $G’$ provides little reduction in the neighbor set $N(M[u’])$. Instead, it introduces substantial memory overhead, as $G’$ must maintain a candidate set for every query vertex. As a result, EGSM faces both high memory consumption and construction-time overhead. Given these limitations, STMatch and T-DFS avoid building $G’$ and directly enumerate on the original graph $G$. However, the large $d_{\max}$ in $G$ causes the execution stack to grow significantly, leading to high memory overhead. Consequently, existing GPU-based methods suffer from severe scalability and parallel efficiency bottlenecks on this workload. Moreover, as the graph follows the power-law degree, the warp can be also frequently underutilized.

In summary, although T-DFS, STMatch, and EGSM all employ coarse-grained parallel execution and can run on the two types of workloads, each is optimized for a distinct workload category. In contrast, gMatch is designed to efficiently support both workload types. Our focus is on accelerating the parallel search phase, which is the core computation shared across both scenarios.

\FloatBarrier
\end{document}